\documentclass[twocolumn,showpacs,amsmath,amssymb,prd]{revtex4}
\usepackage{graphicx}
\def\ba{\begin{eqnarray}}
\def\ea{\end{eqnarray}}

\newcommand{\la}{\stackrel{<}{ _{\sim}}}
\newcommand{\ga}{\stackrel{>}{ _{\sim}}}

\usepackage{graphicx}% Include figure files
\usepackage{dcolumn}% Align table columns on decimal point
\usepackage{bm}% bold math
\usepackage{epsf}

\begin{document}

\newcommand{\geta}{\overline{\eta}}

\title{General relativistic hydrodynamics with viscosity: contraction,
catastrophic collapse, and disk formation in hypermassive neutron stars}

\author{Matthew D. Duez}

\author{Yuk Tung Liu}

\author{Stuart L.\ Shapiro}
\altaffiliation{Department of Astronomy \& NCSA, University of Illinois
at Urbana-Champaign, Urbana, IL 61801}

\author{Branson C. Stephens}

\affiliation{Department of Physics, University of Illinois at
Urbana-Champaign, Urbana, IL~61801}

\begin{abstract}
Viscosity and magnetic fields drive differentially rotating stars toward uniform 
rotation, and this process has important consequences in many 
astrophysical contexts.  For example, merging binary neutron stars 
can form a ``hypermassive'' remnant, i.e.\ a differentially 
rotating star with a mass greater than would be possible for a uniformly 
rotating star.  The removal of the centrifugal support provided by differential 
rotation can lead to 
delayed collapse of the remnant to a black hole, accompanied by a delayed 
burst of gravitational radiation.  Both magnetic fields and 
viscosity 
alter the structure of differentially rotating stars on secular 
timescales, and tracking this evolution presents a strenuous challenge to numerical 
hydrodynamic codes.  Here, we present the first evolutions of rapidly rotating 
stars with shear viscosity in full general relativity.  We
self-consistently include viscosity in our relativistic 
hydrodynamic code by solving the fully relativistic Navier-Stokes equations.  
We perform these calculations both in axisymmetry and in full 3+1 dimensions.
In axisymmetry, the resulting reduction in computational costs allows 
us to follow secular evolution with high resolution over dozens of rotation periods
(thousands of $M$).  
We find that viscosity operating in a hypermassive star generically
leads to the formation of a compact, uniformly rotating core
surrounded by a low-density disk.  These uniformly rotating cores are 
often unstable to gravitational collapse.  We follow the collapse in such cases
and determine the mass and the spin of the final black hole and ambient disk. 
However, viscous braking of differential rotation in hypermassive 
neutron stars does not always lead to catastrophic collapse, especially 
when viscous heating is substantial.  The stabilizing influences of
viscous heating, which generates enhanced thermal pressure, and centrifugal support
prevent collapse in some cases, at least until 
the star cools.  In all cases studied, the rest mass of the resulting disk is 
found to be 10-20\% of the original star, whether surrounding
a uniformly rotating core or a rotating black hole. 
This study represents an important step toward understanding 
secular effects in relativistic stars and foreshadows more detailed, future 
simulations, including those involving magnetic fields.

\end{abstract}


\maketitle

\section{Introduction}
\label{intro}

The field of numerical relativity has matured to a stage where it is 
possible to simulate realistic systems of astrophysical interest. 
In this paper, we 
examine the global effects of viscosity on differentially rotating, 
relativistic stars. Viscosity can have significant 
effects on the stability of neutron stars. For example, it can 
drive a secular bar instability in rapidly rotating neutron stars, as 
shown in Newtonian gravitation~\cite{c69,sl96} and in 
general relativity~\cite{sz98}. Viscosity can
suppress the $r$-modes~\cite{lom98,jjl0102} and other gravitational-radiation 
driven instabilities, including the secular bar modes~\cite{ll7783}. 
Viscosity also destroys differential 
rotation, and this can cause significant changes in the 
structure and evolution of differentially rotating massive 
neutron stars.

Differentially rotating neutron stars can support significantly more rest mass
than their nonrotating or uniformly rotating counterparts, making ``hypermassive'' 
neutron stars possible~\cite{bss00,lbs03}. Such hypermassive neutron stars can 
form from the coalescence of neutron star 
binaries~\cite{rs99,Shibata:1999wm,stu03} or from rotating core 
collapse. 
The stabilization arising from differential rotation, although 
expected to last for many dynamical timescales, will ultimately be destroyed 
by magnetic braking and/or viscosity~\cite{bss00,s00}. 
These processes drive the star to uniform rotation, which cannot 
support the full mass of the hypermassive remnant.  
This process can lead to ``delayed'' catastrophic collapse to a black hole, 
possibly accompanied by some mass loss. Such a delayed 
collapse might emit a delayed gravitational wave signal detectable by 
laser interferometers. Moreover, the collapse, together with 
any residual gas in an ambient accretion disk, could be the origin of a 
gamma-ray burst (GRB).

Both magnetic fields and viscosity can destroy differential rotation in 
a rapidly rotating star~\cite{s00,css03,ls03}. Simple estimates 
show that the magnetic braking (Alfv\'en) timescale for a laminar field
is much shorter than the 
timescale of molecular (neutron) viscosity in a typical massive neutron star. Hence 
magnetic fields are expected to be the principal mechanism driving neutron 
stars toward rigid rotation. 
Phase mixing arising from magnetic braking~\cite{spruit99,ls03}, 
or other possible magnetohydrodynamic instabilities~\cite{spruit99,balbus98} 
might stir up turbulence. Turbulent shear viscosity could then dominate 
the subsequent evolution.  In this paper, we are primarily interested in
identifying the global evolutionary consequences of shear viscosity in a
relativistic star, independent of the detailed nature or origin of the
viscosity.

To explore the consequences of the loss of differential rotation in 
equilibrium stars, we study the secular evolution of 
differentially rotating relativistic stars in the presence of a shear viscosity. 
Viscosity and magnetic fields have two things in common: (1) they both change 
the angular velocity profiles of a differentially rotating star, 
and (2) they both act on {\em secular} timescales, which can be 
many rotation periods. The latter inequality poses a severe
challenge to numerical simulations using a hydrodynamic code. 
It is too taxing for a {\it hydrodynamic} code 
using an explicit differencing scheme to evolve a star for 
physical realistic secular timescales. 
To solve this problem, we artificially 
amplify the strength of viscosity so that the viscous timescale 
is short enough for numerical treatment. However, we keep the viscous 
timescale substantially longer than 
the dynamical timescale of the stars, so that the evolution of the star remains 
quasi-stationary. We then check the validity of our results  
by reducing the viscosity on successive runs and 
testing that the viscosity-induced physical behavior is unchanged; rather, 
only the timescale changes and does so inversely with
the strength of viscosity.
A more detailed discussion of the expected scaling is presented in 
Section~\ref{justification}~\cite{footnote0}.

To study viscous evolution, we need to perform long simulations 
in full general relativity. Typically, 
we evolve the stars in axisymmetry. This allows us to follow the 
secular evolution of the stars with high resolution in a reasonable 
amount of time. Viscosity can, however, drive nonaxisymmetric 
instabilities when a star is rapidly rotating. To test for such 
instabilities, we also perform lower-resolution, three-dimensional (3D)
simulations on the most rapidly rotating stars we consider.

For non-hypermassive neutron stars that are slowly and differentially 
rotating, we find that viscosity simply drives the whole star to rigid 
rotation. If the non-hypermassive neutron star is rapidly and 
differentially rotating, however, viscosity drives the inner core to 
rigid rotation and, at the same time, expels the material in the outer 
layers. The final system in this case consists of a rigidly-rotating 
core surrounded by a low-density, ambient disk in quasi-stationary 
equilibrium.

Our most interesting results concern the fate of hypermassive 
neutron stars. We numerically evolve four models with 
different masses and angular momenta. We find that in all cases, 
viscosity drives the cores to rigid rotation and transports angular
momentum outwards into the envelope. As a result, the core 
contracts in a quasi-stationary manner, and the outer layers expand to
form a differentially rotating torus.  Of the 
four models we have studied, the star with the highest mass collapses 
to a black hole, with about 20\% of the rest mass leftover to form a massive 
accretion disk. On the contrary, the other three stars do not collapse to 
black holes, but form star + disk systems, similar to the final state of 
the rapidly rotating
non-hypermassive neutron stars described above. As will be discussed in 
Section~\ref{rad_cooling}, viscosity generates heat so that the stars 
do not evolve adiabatically in general. The extra thermal pressure 
due to viscous heating helps to support the stars. We also consider 
the limit of rapid cooling, whereby the heat generated by viscosity is 
immediately removed from the stars. Of 
the three stars which do not collapse to black holes in the no-cooling 
limit, we found that 
the one with the lowest angular momentum undergoes catastrophic 
collapse in the rapid-cooling limit. About 
10\% of the rest mass is leftover to form an accretion disk
in this case. To test the validity of the axisymmetric results, 
we perform 3D simulations to check for any nonaxisymmetric instabilities. 
We do not find any unstable nonaxisymmetric modes and the 3D results 
agree with the axisymmetric results.

Our results suggest that viscous braking of differential rotation in a
hypermassive neutron star can, but does not always, lead to 
catastrophic collapse. When catastrophic collapse does occur, 
the remnant is a black hole surrounded by a massive accretion disk. 
This outcome is very different from that of the collapse of an 
unstable, rigidly-rotating ``supramassive'' neutron star, in which 
the whole star collapses to a black hole, 
leaving only a tiny amount of material to form a disk~\cite{sbs00,s03}. 
Many models for GRBs require a massive disk around
a rotating black hole to 
supply energy by neutrino processes~\cite{rosswog}. 
Our results suggest that viscous forces in a hypermassive star 
could lead to the formation of a massive disk around such a black hole.

The structure of this paper is as follows. In Section~\ref{setup}, 
we derive the relativistic Navier-Stokes equations containing shear
viscosity in a 3+1 form suitable for numerical integration, and 
describe how we evolve them in both axisymmetry and 
full 3+1 dimensions. We then describe in Section~III several tests
that we perform to check our code.  We present the results of our
simulations on five selected stars in Section~IV. Finally, we briefly
summarize and discuss our conclusions in Section~V.

\section{Formalism and Numerical Methods}
\label{setup}

\subsection{Evolution of the gravitational fields}

Throughout this paper, Latin indices denote spatial components
(1-3) and Greek indices denote spacetime components (0-3). 
We adopt geometrized units, so that $G = c = 1$.
We evolve the 3-metric $\gamma_{ij}$ and the extrinsic curvature $K_{ij}$
using the BSSN formulation~\cite{bssn}.  The fundamental variables for
BSSN evolution are
\begin{eqnarray}
  \phi &\equiv& {1\over 12}\ln[\det(\gamma_{ij})]\ , \\
  \tilde\gamma_{ij} &\equiv& e^{-4\phi}\gamma_{ij}\ , \\
  K &\equiv& \gamma^{ij}K_{ij}\ , \\
  \tilde A_{ij} &\equiv& e^{-4\phi}(K_{ij} - {1\over 3}\gamma_{ij}K)\ , \\
  \tilde\Gamma^i &\equiv& -\tilde\gamma^{ij}{}_{,j}\ .
\end{eqnarray}
The evolution and constraint equations for
these fields are summarized in~\cite{paperI} (hereafter Paper~I).  In the
presence of matter, these evolution equations contain the following source
terms:
\begin{eqnarray}
  \rho &=& n_{\alpha}n_{\beta}T^{\alpha\beta}\ , \nonumber \\
  S_i  &=& -\gamma_{i\alpha}n_{\beta}T^{\alpha\beta}\ , \label{source_def} \\
  S_{ij} &=& \gamma_{i\alpha}\gamma_{j\beta}T^{\alpha\beta}\ ,\nonumber
\end{eqnarray}
where $T^{\alpha\beta}$ is the stress tensor, and
$n_{\alpha} = (-\alpha,0,0,0)$ is the future-directed unit normal to 
the time slice.
One must impose gauge conditions which specify the lapse $\alpha$
and the shift $\beta^i$.  We use a K-driver lapse and Gamma-driver
shift, as described in Paper~I.  The numerical implementation of the equations
is discussed in Paper~I, with some improvements to enhance
stability described in~\cite{dsy03}.  The latter are particularly relevant
for the post-collapse version of our code that we implement with black-hole
excision.

\subsection{3+1 relativistic Navier-Stokes equations}

We treat the matter in our neutron stars as an imperfect fluid
with a shear viscosity, but no bulk viscosity and no heat conduction. 
The stress tensor for the fluid is
\begin{equation}
  \label{Tuv}
  T_{\mu\nu} = (\rho_0 + \rho_0\epsilon + P)u_{\mu}u_{\nu} + P g_{\mu\nu}
  - 2\eta \sigma_{\mu\nu}\ .
\end{equation}
Here, $\rho_0$, $\epsilon$, $P$, and $u_{\mu}$ are the rest-mass
density, specific internal energy, pressure, and fluid four-velocity,
respectively.  The quantity $\eta$ is the coefficient of viscosity and is related
to the kinematic viscosity $\nu$ by $\eta = \rho_0\nu$.  The
shear tensor $\sigma_{\mu\nu}$ is defined by
\cite{mtw}
\begin{equation}
\label{sigma_def}
  \sigma_{\mu\nu} \equiv u_{(\mu;\nu)} + a_{(\mu}u_{\nu)} 
  - {1\over 3}u^{\alpha}{}_{;\alpha}
  (g_{\mu\nu} + u_{\mu}u_{\nu})\ ,
\end{equation}
where $a^{\mu}$ is the fluid 4-acceleration.  We assume a $\Gamma$-law
equation of state
\begin{equation}
  \label{ideal_P}
  P = (\Gamma - 1)\rho_0\epsilon\ .
\end{equation}
Our fundamental fluid variables are
\begin{eqnarray}
  \rho_{\star} &\equiv& \rho_0\alpha u^0e^{6\phi}\ , \\
  e_{\star} &\equiv& (\rho_0\epsilon)^{1/\Gamma}\alpha u^0e^{6\phi}\ , \\
  \tilde S_k &\equiv& \rho_{\star}hu_k\ ,
\end{eqnarray}
where $h = 1+\epsilon+P/\rho_0$ is the specific enthalpy. 
The conservation of stress-energy
\begin{equation}
  \label{eom}
  T^{\mu\nu}{}_{;\nu} = 0
\end{equation} 
and the law of baryon number conservation
\begin{equation}
  \label{barycons}
  \nabla_{\mu}(\rho_0 u^{\mu}) = 0
\end{equation}
give the relativistic continuity, energy, and Navier-Stokes equations
\begin{eqnarray}
  \label{evolve_rhostar}
  \partial_t\rho_{\star} &+& \partial_i(\rho_{\star}v^i) \ \, =\ 0 \\
  \label{evolve_estar}
  \partial_t e_{\star} &+& \partial_i(e_{\star}v^i) \, \ =\ 
	  {2\over\Gamma}\alpha e^{6\phi}\eta
	  (\rho_0\epsilon)^{(1-\Gamma)/\Gamma}\sigma^{\alpha\beta}\sigma_{\alpha\beta} \ \ \ \\
%	  & & \qquad \ \ \ \ \ \ \ \ \ \ \
%	  \times \sigma^{\alpha\beta}\sigma_{\alpha\beta} \nonumber \\
	  \label{evolve_stilde}
	  \partial_t\tilde S_k &+& \partial_i(\tilde S_k v^i) \ =\ 
	  -\alpha e^{6\phi}P_{,k}
	  + 2(\alpha e^{6\phi}\eta\sigma_k^{\mu})_{,\mu}\ \  \\
	  &+& \alpha e^{6\phi}g^{\alpha\beta}{}_{,k}
	  \left(\eta\sigma_{\alpha\beta}
	  - {1\over 2}\rho_0hu_{\alpha}u_{\beta}\right)\ ,
%	  & & ~~~~~~~~~~~~~~~ - {1\over 2}\rho_0hu_{\alpha}u_{\beta})\ ,
	  \nonumber
\end{eqnarray}
where $v^i = u^i/u^0$ is the 3-velocity.  The
quantity $u^0$ is determined by the normalization condition
$u^{\nu}u_{\nu} = -1$, which yields
\begin{equation}
  \label{w}
  \displaystyle w^2 = \rho_{\star}^2 
  + e^{-4\phi}\tilde\gamma^{ij}\tilde S_i \tilde S_j
  \left[1 + {\Gamma e_{\star}{}^{\Gamma}\over
      \rho_{\star}(we^{6\phi}/\rho_{\star})^{\Gamma-1}}\right]^{-2}\ ,
\end{equation}
where $w = \rho_{\star}\alpha u^0$.

The stress-tensor $T^{\mu\nu}$ generates the following source
terms in the field evolution equations:
\begin{eqnarray}
\label{rho_source}
\rho &=& h w e^{-6\phi} - P \\
     & &   - {2\eta\over\alpha^2}(\sigma_{tt}-2\sigma_{ti}\beta^i
	                       + \sigma_{ij}\beta^i\beta^j) \ ,
			       \nonumber \\
\label{Si_source}
S_i &=& e^{-6\phi} \tilde S_i
        - {2\eta\over\alpha}(\sigma_{ti} - \sigma_{ij}\beta^j) \ , \\
\label{Sij_source}
\displaystyle S_{ij} &=& {e^{-6\phi}\over w h}\tilde S_i \tilde S_j
                          + P \gamma_{ij} - 2\eta\sigma_{ij} \ .
\end{eqnarray}

\subsection{2+1 relativistic Navier-Stokes equations}

Many of the systems we evolve possess and maintain symmetry about
their rotation axis, which we set to be the $z$-axis.  Then we can
eliminate one dimension and simplify the equations.  We
utilize axisymmetry and follow~\cite{abbhstt01,Shibata:2000} to evolve the field
and hydrodynamic variables on the $y = 0$ plane.  The data off of
this plane can be obtained by rotating the data on this plane.  As
we explain in Section~\ref{axisym_methods}, we find it advantageous when
performing 2+1 simulations to evolve
the hydrodynamic equations (\ref{evolve_rhostar})-(\ref{evolve_stilde})
in cylindrical coordinates but on a Cartesian ($xz$) grid.  On the
$y = 0$ plane, the cylindrical
coordinates $\varpi = \sqrt{x^2 + y^2}$, $z$, and $\varphi = \arctan(y/x)$
are related to the Cartesian coordinates $x$, $y$, and $z$ as follows:
$\varpi\leftrightarrow x$, $\varpi\varphi\leftrightarrow y$,
$z\leftrightarrow z$.  Using these relations, Eqs.~(\ref{evolve_rhostar})-(\ref{evolve_stilde}) in 
cylindrical coordinates can be written
\begin{eqnarray}
  &&\partial_t \rho_{\star} + {1 \over x}\partial_B(\rho_{\star}x v^B)=0
  \label{continuity}\\
  &&\partial_t (\tilde S_A-2\alpha e^{6\phi}\eta\sigma^0_A)
  + {1 \over x}\partial_B(x \tilde S_A v^B - 2x\alpha e^{6\phi}\eta\sigma^B_A)
  \nonumber \\
  &&\qquad\ \ ={1\over x}(\tilde S_y v^y - 2x\alpha e^{6\phi}\eta\sigma^y_y)\delta_{Ax}
  -\alpha e^{6\phi}\partial_A P \label{eulerA} \\
  && \qquad\ \ \ \ + \alpha e^{6\phi}g^{\alpha\beta}{}_{,A}
  \left[-{1\over 2}\rho_0hu_{\alpha}u_{\beta} + \eta\sigma_{\alpha\beta}
    \right] \nonumber\\
  &&\partial_t (\tilde S_y - 2\alpha e^{6\phi}\eta\sigma^0_y)
  \label{eulery}\\
  && \qquad\ \ + {1 \over x^2}\partial_B(x^2 \tilde S_y v^B - 2\alpha e^{6\phi}\eta x^2 \sigma^A_y)=0 \nonumber \\
  && \partial_t e_{\star} + {1 \over x}\partial_B(e_{\star} x v^B)  \label{energy} \\
  && \qquad\ \  = {2\over\Gamma} \alpha e^{6\phi}\eta
  (\rho_0\epsilon)^{(1-\Gamma)/\Gamma}\sigma^{\alpha\beta}\sigma_{\alpha\beta} \ ,
  \nonumber
\end{eqnarray}
where the indices $A$ and $B$ run over $x$ and $z$
(c.f. Eq.(2.10)-(2.13) of~\cite{Shibata:2000}).

\subsection{Hierarchy of timescales}
\label{timescales}

There are two dynamical timescales for a rotating star.  Gravity
provides the free-fall timescale $\tau_{\rm FF}$
\begin{equation}
  \tau_{\rm FF} \sim \left( \frac{R^3}{M}\right)^{1/2} \sim 
10^{-4} \left({M/R\over 0.2}\right)^{-3/2}
  \left({M\over 2 M_{\odot}}\right) {\rm s}\ ,
\end{equation}
where $M$ is the gravitational mass of the star (or merged binary
remnant) and $R$ is the radius.  If the star
is rotating, its rotation period $P_{\rm rot}$ provides another
important timescale:
\begin{equation}
  P_{\rm rot} = {2\pi\over\Omega} = 1.6 \times 10^{-3}
  \left({\Omega\over 4000 {\rm s}^{-1}}\right)^{-1} {\rm s} \ .
\end{equation}
Dynamical instabilities (e.g. instability to radial collapse or
to dynamical bar formation) will act on the above timescales.

The stars we study are dynamically stable initially, so their 
structure is altered on secular timescales. Rotating 
compact stars may be secularly unstable to gravitational-radiation 
driven instabilities. The strongest instabilities of this kind are 
the (nonaxisymmetric) $r$-modes and the bar mode.
The timescale of the $l=m=2$ $r$-mode instability is given by~\cite{lom98}
\begin{equation}
  \tau_r^{\rm GW} \sim 50 \left( \frac{\Omega}{4000 {\rm s}^{-1}}
\right)^{-6} \left( \frac{M/R}{0.2}\right)^4 \left( \frac{M}{2M_{\odot}} 
\right)^{-5} \, {\rm s} \ .
\end{equation}
The gravitational-radiation driven (Dedekind) bar-mode instability occurs if the 
star is rapidly rotating so that $T/|W| > \beta_s$, where $T/|W|$
is the ratio of kinetic to gravitational potential energy . The threshold  
$\beta_s \approx 0.14$ for a Newtonian star ($M/R \ll 1$), and {\it decreases} 
as the compactness of the star (i.e.\ $M/R$) increases~\cite{secbar}. 
The timescale of this instability is estimated to be~\cite{fs7578}
\begin{equation}
  \tau_{\rm bar}^{\rm GW} \sim 0.1 \left( \frac{M/R}{0.2}\right)^{-4} 
\left( \frac{M}{2M_{\odot}}\right)
\left( \frac{T/|W|-\beta_s}{0.1} 
\right)^{-5} \, {\rm s} \ . 
\end{equation}
Viscosity alone can also drive a (Jacobi) bar-mode instability.  The threshold
is identical for a Newtonian star ($\beta_s \approx 0.14$) but {\it increases}
as the compaction increases~\cite{sz98}.  The relevant timescale is
(see \cite{c69},~p.99)
\begin{equation}
  \tau_{\rm bar}^{\rm vis} \sim {R^2\over 4\nu}
  \left( \frac{T/|W|-\beta_s}{0.1} \right)^{-1} \, {\rm s} \ . 
\end{equation}
This is comparable to the viscous timescale 
$\tau_{\rm vis}\sim R^2/\nu$ discussed below.

Magnetic fields coupled to the matter will redistribute
angular momentum.  In fact, even an initially small magnetic field
frozen into the matter will be wound up and can destroy differential rotation in the
star on an Alfv\'en timescale~\cite{spruit99,s00,bss00}:
\begin{equation}
  \tau_{\rm B} \sim 100 \left({B\over 10^{12}G}\right)^{-1}
  \left({M/R\over 0.2}\right)^{1/2} {\rm s}\ .
\end{equation}

Viscosity will also
redistribute angular momentum on a viscous timescale $\tau_{\rm vis}$. 
One form of viscosity present in neutron stars is due to the transport 
of energy and momentum of neutrons. 
This viscosity acts on a timescale~\cite{fi79,cl87}
\begin{equation}
  \tau_{n,\rm vis} \sim 10^8 T_9^2
  \left({M\over 2 M_{\odot}}\right)^{9/2}
  \left({M/R\over 0.2}\right)^{-23/4} {\rm s}\ ,
\end{equation}
where $T_9 = T/10^9 K$, and $T$ is the characteristic temperature.

It is widely believed that, for cold neutron stars ($T \la 10^9 K$),
the neutron fluid in the inner crust condenses into a superfluid
of ${}^1S_0$ Cooper
pairs~\cite{ccdk93}, while in the interior, the neutrons could form
a ${}^3P_2$ superfluid~\cite{hgrr70} (although this is less certain),
and the protons a ${}^1S_0$ superfluid.  In the case of neutron
superfluidity, $\tau_{n,\rm vis}$
will vanish, and the dominant viscosity will be due to electron-electron
scattering~\cite{fi76,cl87}
\begin{equation}
  \tau_{e,\rm vis} \sim 10^8 T_9^2
  \left({M\over 2 M_{\odot}}\right)^{4}
  \left({M/R\over 0.2}\right)^{-5} {\rm s}\ .
\end{equation}
Electron and proton fluids are forced to move together in the
MHD limit~\cite{m98}. 
Differences in velocity between the neutron and proton-electron
fluids are damped fairly quickly by mutual friction~\cite{als84,m98}. 

Viscosity can be used as a model for turbulence in certain 
situations. Turbulence may occur in young neutron stars as a result of
pure hydrodynamic effects or magnetic instabilities~\cite{balbus98}. 
Turbulence 
is often modeled by the ``$\alpha$-disk'' law, 
in which a shear stress $T_{\varpi\varphi}=-\alpha P$ is added to the 
hydrodynamic equation (see e.g.~\cite{st83}, Chap.~14). 
Here $\alpha$ is a nondimensional constant (which should not 
be confused with the lapse function) with values in the range 
$0.001 \lesssim \alpha \lesssim 1$.  The viscosity in this model roughly
corresponds to 
$\nu \sim l_{\rm turb} v_{\rm turb} \sim\alpha Rc_{\rm s}$, where
$v_{\rm turb}$ is the velocity of turbulent cells relative to
the mean fluid motion, $l_{\rm turb}$ is the size of the
largest turbulent cell, and $c_s$ is the sound speed.
The corresponding timescale is 
\begin{equation}
  \tau_{\rm vis}^{\rm turb} \sim \frac{1}{\alpha} \tau_{\rm FF} 
\sim \frac{10^{-4}}{\alpha} \left( \frac{M/R}{0.2} \right)^{-3/2} 
\left( \frac{M}{2M_{\odot}} \right) \, {\rm s} \ ,
\label{t_turb}
\end{equation}
which is much shorter than 
all the other secular timescales. Hence turbulent viscosity, if 
present, is likely to dominate the secular evolution of differentially 
rotating stars. 

Thermal energy is radiated away primarily by neutrinos.  For hot
neutron stars ($T \ga 10^9 K$), the cooling is dominated by the
direct URCA process, and the star cools on a timescale
$\tau_{\rm cool} \sim 10^2 T_9^{-4}$~s (see \cite{p92} and \cite{st83}, Chap. 11).  
For cooler neutron
stars, the cooling is dominated by the modified URCA process, and
the star cools on a timescale $\tau_{\rm cool} \sim 10^7 T_9^{-6}$~s~\cite{p92}. 
Depending on the temperature and the nature of the viscosity, the cooling
timescale may be greater than or less than the viscous timescale.  If
$\tau_{\rm vis}\ll\tau_{\rm cool}$, then the heat generated by viscosity will
build up inside the star.  Otherwise, it will be radiated away as quickly
as it is generated.  We study both limits in this paper.

\subsection{Dynamically modeling secular effects}
\label{justification}

Secular effects will in general take many rotation periods to significantly
affect the structure or velocity profile of a differentially rotating
star. This poses a challenge to the numerical treatment of these changes. 
Because of the short Courant timestep required for numerical stability, 
it would
be computationally prohibitive to evolve a star for such a long time
using an explicit finite differencing scheme. 
The use of a fully implicit scheme for the finite differencing can
allow stable evolutions with larger $\Delta T$.  Each timestep is,
however, much more computationally expensive as it involves matrix
inversion.  Moreover, no fully 
implicit routine for the coupled Einstein field and relativistic 
hydrodynamic equations exists at present.

The secular timescales are so much longer than the dynamical timescales
that the star can be thought of as evolving quasi-statically.
Therefore, it might be possible to treat the secular evolution 
in the quasistatic 
approximation, as in typical stellar evolution (Henyey) codes, 
by constructing a
sequence of equilibrium configurations up to the moment that stable 
equilibrium can no longer be sustained. This approach has been used 
to study the viscous evolution of differentially rotating 
white dwarfs~\cite{durisen}.  However, building the required equilibrium
models in full general relativity is a nontrivial task. It would be 
particularly difficult to identify the meridional currents and 
core-halo bifurcation that often arise in rapidly rotating 
configurations.  More significantly, it would not be possible to
follow the evolution of the configuration with a quasistationary
approach if a dynamical instability
(i.e. collapse) is triggered during the secular evolution.

Instead, we use our relativistic hydrodynamic code and artificially 
amplify the strength of viscosity so that the viscous timescale 
is short enough to make numerical treatment tractable.  However, we
keep the viscous timescale sufficiently long that the hierarchy of
timescales is maintained, and the secular evolution still proceeds in a
quasi-stationary manner. 
The behavior of the real system can then be determined
by rescaling the time variable to adjust the viscous timescale to its
physical value. The characteristic viscous timescale is 
\begin{equation}
\label{tvis_general}
\tau_{\rm vis} \sim \rho R^2 <\!\eta\!>^{-1}\ ,
\end{equation}
where $<\!\!\eta\!\!>$ is an averaged value of $\eta$ across the star. 
Suppose we evolve the same star, once with $\tau_{\rm vis} = \tau_1$ and
once with $\tau_{\rm vis} = \tau_2$.  If both $\tau_1$ and $\tau_2$ are
large enough so that they do not compete with the dynamical timescale,
but shorter than any other secular timescale
(see, e.g. Section~\ref{timescales}), then the configuration of the
star with viscosity $\tau_1$ at time $t$ will be the same as the
configuration of the star with viscosity $\tau_2$ at time
$(\tau_2/\tau_1) t$. By varying $\tau_{\rm vis}$ over a wide 
range and corroborating this scaling, we are confident that 
the physical behavior we observe is real. We can then scale the 
result of numerical simulations to the appropriate strength of 
viscosity, provided that the physically relevant viscous 
timescale is much shorter than all the other secular timescales
(e.g., $\tau_r^{\rm GW}$, $\tau_{\rm bar}^{\rm GW}$). 

As discussed in the previous subsection, turbulent viscosity is 
likely to dominate the secular evolution. 
%The simple $\alpha$-disk 
%model is not suitable for the fluid in a star. This is because 
%the additional stress tensor $T_{\varpi\varphi}=-\alpha P$ has the 
%undesirable feature that it will change the velocity profile 
%even if the star is rigidly rotating. Instead, 
We adopt the stress 
tensor as in Eq.~(\ref{Tuv}) and specify the viscosity $\eta$ suitable 
for modeling turbulence. 
We consider the turbulent viscosity described in~\cite{ll87}:
\begin{equation}
\eta \sim \rho l_{\rm turb} v_{\rm turb} \ .
\end{equation}
(This viscosity law is also used in 
some accretion-disk models~\cite{balbus98}.)
Typically, $v_{\rm turb}$ is proportional to the sound speed $c_s$. Hence 
$\eta \sim l_{\rm turb} \rho \sqrt{P/\rho} \sim (l_{\rm turb}/c_s) P$.
For simplicity, we assume that $l_{\rm turb}/c_s$ is constant inside 
the star. Then we have 
\begin{equation}
  \eta = \nu_P P\ ,
\end{equation}
where $\nu_{\rm P}$ is a constant, and is related to the coefficient of 
kinematic viscosity $\nu$ by $\nu_P = (\rho_0/P) \nu$.  In our numerical
simulations we specify the value of $\nu_P$ for each run.

We model the initial stars as rotating polytropes with polytropic index $n=1$, so
that $P = \kappa \rho_0^2$. It is convenient to rescale all quantities 
with respect to $\kappa$. Since $\kappa^{1/2}$ has dimensions of length, 
we can define the following nondimensional variables~\cite{cst92}
\begin{eqnarray}
  \bar{x}^{\mu} = \kappa^{-1/2} x^{\mu} \ , & 
  \bar{\Omega} = \kappa^{1/2} \Omega \ , & \\ 
  \bar{M}=\kappa^{-1/2} M \ , &
  \bar{R} = \kappa^{-1/2} R \ , & \\ 
  \bar{\rho}_0 = \kappa \rho_0 \ \ \ \ \ \ \ , & 
  \bar{\nu}_P = \kappa^{-1/2} \nu_P \ , &
\end{eqnarray}
where the spacetime coordinates are $x^{\mu}=(t,x,y,z)$.
However, to simplify our notation, we will drop all the bars. Hereafter, 
all variables are understood to be in ``$\kappa=1$ units.''

Using Eq.~(\ref{tvis_general}), we can see
that $\tau_{\rm vis}$ scales with $R$, $\rho$, and $\nu_P$ as
\begin{equation}
  \tau_{\rm vis} \sim {R^2\over\nu} \sim {R^2\rho\over\nu_PP}
\end{equation}
which for $n = 1$ becomes
\begin{equation}
  \tau_{\rm vis} \sim {R^2\over \rho \nu_P} \ .
\end{equation}
For definiteness, we take $\tau_{\rm vis}$ to be
\begin{equation}
\label{tvis_P}
\tau_{\rm vis} = \lambda {R_{\rm eq}^2\over \rho_{0,\max}\nu_P}\ ,
\end{equation}
where $R_{\rm eq}$ is the equatorial radius, $\rho_{0,\max}$ is
the maximum value of $\rho_0$ in the star, and $\lambda$ is a
dimensionless constant.  We use the constant $\lambda$
to approximately match $\tau_{\rm vis}$ to the rate of decay of
differential rotation observed by our simulations.  With the
appropriate $\tau_{\rm vis}$, the value of $\sigma_{\mu\nu} \sigma^{\mu \nu}$, 
which is proportional to the rate of energy dissipation 
[see Eq.~(\ref{evolve_estar})], is expected to decay like
\begin{equation}
\label{sigma_falloff}
\sigma_{\mu\nu} \sigma^{\mu \nu} \approx (\sigma_{\mu\nu} \sigma^{\mu \nu})|_{t=0}
\exp(-2t/\tau_{\rm vis})\ .
\end{equation}
We measure $\tau_{\rm vis}$ by numerically evolving a given star, and
observing the decay
with time of $<\!\sigma^{\mu\nu}\sigma_{\mu\nu}\!>$, the average value
of $\sigma^{\mu\nu}\sigma_{\mu\nu}$ throughout the star, weighted
by rest density. We determine the value of $\lambda$ by requiring that 
$\tau_{\rm vis}$ roughly corresponds to the e-folding time of the decay 
of $\sqrt{<\!\sigma^{\mu\nu}\sigma_{\mu\nu}\!>}$. 
Carrying out this measurement on a sampling of 
the stars used below, we find that $\lambda \approx 0.23$ in all cases. 
Thus we use $\lambda=0.23$ to define $\tau_{\rm vis}$ in the sections below.

We note that the turbulent viscosity adopted above for our numerical treatment
is roughly equivalent to an ``$\alpha$'' model provided we identify
$\alpha\sim(c_{\rm s}/R)\nu_P\sim(M/R^3)^{1/2}\nu_P$.  Eq.~(\ref{tvis_P})
for $\tau_{\rm vis}$ is then equivalent to Eq.~(\ref{t_turb}).

\subsection{Radiative cooling}
\label{rad_cooling}
Our stars do not evolve isentropically.  Viscosity heats the matter
on a timescale $\tau_{\rm vis}$, as
shown by Eq.~(\ref{evolve_estar}).  At the same time, neutrino radiation
carries away heat, cooling the star on a timescale $\tau_{\rm cool}$.    
We will carry out simulations below in two opposite limits, which we describe in
some detail in Appendix~\ref{cooling_app}.  In the
{\it no-cooling} limit, $\tau_{\rm cool} \gg \tau_{\rm vis}$, so we ignore
radiative cooling and simply evolve
Eqs.~(\ref{evolve_rhostar})-(\ref{evolve_stilde}).  In the {\it rapid-cooling}
limit, $\tau_{\rm cool} \ll \tau_{\rm vis}$, so we 
evolve Eqs.~(\ref{evolve_rhostar})-(\ref{evolve_stilde})
as before, but {\it without including} the viscous heating term in
the energy equation [Eq.~(\ref{evolve_estar})].  
This will allow net heating by adiabatic compression but not by viscosity. 
The viscous heat is assumed to be (instantaneously) lost by radiation in this
limit, while viscous braking proceeds.  The emitted radiation will carry off
some momentum as well as energy, causing a
modification of Eq.~(\ref{evolve_stilde}), but this will have a much
smaller effect provided the 
luminosity does not exceed the (neutrino) Eddington luminosity. Baumgarte and
Shapiro~\cite{bs98a} investigated the loss of angular momentum in binary
neutron star merger remnants due to radiation, and they found it to
be fairly small.  We therefore feel justified in ignoring radiative
corrections to Eq.~(\ref{evolve_stilde}).  

\subsection{Numerical Implementation}

\subsubsection{2+1 Dimensional Code}
\label{axisym_methods}

Our hydrodynamical scheme employs Van Leer-type advection with artificial
viscosity to handle shocks.  We also use a ``no-atmosphere'' approach, 
in which the density
at any point on our grid can fall exactly to zero.  Our hydrodynamical 
algorithms are described in detail in Paper~I.  
We have evolved the above equations both in two dimensions, assuming
axisymmetry, and in three dimensions.  Using axisymmetry
saves us computational time and allows us to use higher resolution. 
However, 3D
runs must still be carried out for rapidly rotating systems in order
to check for the occurrence of nonaxisymmetric instabilities.  There
are several ways to evolve in axisymmetry.  One could write the field
and hydrodynamic evolution equations in cylindrical coordinates
($\varpi$,$z$,$\varphi$) and evolve in this coordinate system.  This
has the advantage that one can explicitly remove the dependence of
the variables on $\varphi$.  Unfortunately, there are singularities
in the cylindrical coordinate system which can make the evolution
of the field equations in these coordinates difficult.  Instead, we
choose to
evolve the metric variables ($\tilde\gamma_{ij}$, $\tilde A_{ij}$,
$\phi$, $K$, $\tilde\Gamma^i$, $\alpha$, and $\beta^i$) in axisymmetry
using the Cartoon method~\cite{abbhstt01}.  In this
approach, variables are evolved on a Cartesian grid consisting of three
planes corresponding to $y=-\Delta Y$, $y=0$, and $y=\Delta Y$. 
Then the middle ($y=0$) plane is evolved using the 3D
evolution equations in Cartesian coordinates.  Each time the middle plane
is evolved forward one timestep, the $y=-\Delta Y$ and $y=\Delta Y$ planes
are updated by applying the assumption of
axisymmetry.  Thus, the value of a tensor {\bf f} at location
$(\varpi,z,\pm \varphi)$ on $y = \pm\Delta Y$ is equal to {\bf f} at
$(\varpi,z,0)$ rotated by (coordinate) tensor transformation about the $z$-axis by
angles $\pm \varphi$. 
Since an arbitrary point $(\varpi,z,0)$ will generally
not coincide with any gridpoint on the $y = 0$ plane,
interpolation in $x$ is necessary to apply this update.  We use
third-order polynomial interpolation, so that we do not lose
second-order accuracy.

With the hydrodynamic evolution equations, we also have the choice of
either evolving in cylindrical coordinates or evolving in Cartesian
coordinates using the Cartoon prescription. 
Like Shibata in his work on axisymmetric star
collapse~\cite{Shibata:2000,Shibata:2003}, we choose to evolve the 
fluid variables in
cylindrical coordinates, i.e.\ we use Eqs.~(\ref{continuity})-(\ref{energy}). 
This is superior to using the Cartoon
method because Eq.~(\ref{continuity}) can be finite
differenced in such a way that the total
rest mass will be exactly conserved (except for flow beyond the outer
boundaries). 
In the absence of viscosity, angular momentum also becomes a numerically
conserved quantity.  We have found that evolving the fluid variables in
cylindrical coordinates gives significantly more accurate runs than
evolving via Cartoon hydrodynamics.  The drawback of using 2D
evolutions is
the instability caused by the coordinate singularity on the $x=0$ axis. 
(This instability is also present if we use the 3D Cartesian Navier-Stokes
equations together with the Cartoon boundary condition.)
There are several ways of removing this instability.  
Shibata~\cite{Shibata:2000} adds a small artificial shear viscosity
\begin{equation}
\partial_t(\tilde S_A/\rho_{\star}) = \cdots +
\nu_{\rm art}\rho_{\star}\Delta (\tilde S_A/\rho_{\star})\ ,
\end{equation}
where $\Delta$ is the flat-space Laplacian, and
$\tilde S_i/\rho_{\star} = hu_i$ is the momentum variable evolved by
the code of~\cite{Shibata:2000} (instead of $\tilde S_i$).
 
We have confirmed that adding such a
term to Eq.~(\ref{evolve_stilde}) can stabilize our code.  However,
since we will be studying the effects
of real shear viscosity, we instead choose to remove the instability using
a higher-order dissipation scheme.  Namely, we add a small Kreiss-Oliger
dissipation term~\cite{chlp03}
\begin{equation}
\partial_t\tilde S_A = \cdots
- C_{\rm ko}{(\Delta X\Delta Z)^2\over 16\Delta T} \Delta^2\tilde S_A\ ,
\end{equation}
We use $C_{\rm ko} = 0.2$ for all the simulations reported in this paper.

In both 2D and 3D simulations, we assume that our
system preserves equatorial symmetry across the $z = 0$ plane, and
we therefore only evolve the $z > 0$ portion of the grid.  In
3D runs, we make the added assumption of $\pi$-symmetry,
which allows us to evolve only half of the remaining grid, which we
choose to be the $y > 0$ half.  When performing 2D runs,
we evolve only the region $x > 0$ since the values of variables in the
$x < 0$ region can be deduced from the values on the $x > 0$ region
from the assumed axisymmetry.

\subsubsection{Finite differencing}
We compute all spacial derivatives using standard centered differencing. 
We integrate forward in time using a 3-step iterated Crank-Nicholson
scheme.  So, for example, we evolve the equation $\partial_tf = \dot f(f)$
from timestep $n$ at time $t$ to timestep $n+1$ at time $t+\Delta T$ by
the following algorithm: \\
\texttt{(i) \ Predict: ${}^1f^{n+1} \equiv f^n + \Delta T \dot f(f^n)$}\\
\texttt{(ii) 1st Correct: \\
$~$\hskip40pt  ${}^2f^{n+1} \equiv f^n + \Delta T [0.4\dot f(f^n)+0.6\dot f({}^1f^{n+1})] $}\\
\texttt{(iii) 2nd Correct: \\
$~$\hskip40pt  $f^{n+1} = f^n + \Delta T [0.4\dot f(f^n)+0.6\dot f({}^2f^{n+1})] $} \\
As discussed in Paper I, the coefficients 0.4 and 0.6 were chosen to
improve stability.

In the presence of viscosity, the time-differencing is not
entirely straightforward, due
to the presence of time derivatives of $u_{\mu}$ in $\sigma_{\mu\nu}$,
and of time derivatives of $\sigma_{\mu\nu}$ in the Navier-Stokes
equations.  Thus, $\partial_t\tilde S_k$ (which gives $\partial_tu_k$)
is an expression which itself contains
$\partial_tu_k$ and $\partial^2_tu_k$.  Since the viscosity
is a small perturbing force on the fluid motion, we find that it
is sufficient to split off the viscous terms and integrate them separately
(operator splitting).  In particular, we compute $\partial_tu_k$ and
$\partial_t\sigma^0_k$ appearing in the viscous terms
in a non-time centered way.  Consider the $\tilde S$ evolution equation
for the viscous piece: 
$\partial_t\tilde S = \dot{\tilde S}(\sigma,\dot\sigma)$, where we
have suppressed all indices.  To evolve this equation from timestep
$n$ to timestep $n + 1$, we need to know the time derivatives of $u$
and $\sigma$.  When performing the predictor step, these time derivatives
are approximated by subtracting values of the fields on the timestep $n$
from those on the previous timestep, $n-1$. \\
\texttt{(i) Before predictor step, \\
$~$\hskip20pt  compute $\dot u^{n-1/2} = [u^n - u^{n-1}]/\Delta T$, \\
$~$\hskip40pt  $\sigma^n = \sigma(u^n,\dot u^{n-1/2})$,\\
$~$\hskip40pt  $\dot \sigma^{n-1/2} = [\sigma^n - \sigma^{n-1}]/\Delta T$} \\
Note that these time derivatives are centered at $n -1/2$.  We then carry
out the predictor step.\\
\texttt{(ii) Predict: ${}^1\tilde S^{n+1} = \tilde S^n + \Delta T \dot{\tilde S}(\sigma^n,\dot\sigma^{n-1/2})$} \\
From the predicted values of $u$ and $\sigma$, we construct time derivatives
centered at $n + 1/2$ and use these in the corrector step. \\
\texttt{(iii) compute ${}^1u^{n+1}$ from ${}^1\tilde S^{n+1}$ \\
$~$\hskip40pt  ${}^1\dot u^{n+1/2} = [{}^1u^{n+1} - u^{n}]/\Delta T$, \\
$~$\hskip40pt  ${}^1\sigma^{n+1} = \sigma({}^1u^{n+1},{}^1\dot u^{n+1/2})$,\\
$~$\hskip40pt  ${}^1\dot \sigma^{n+1/2} = [{}^1\sigma^{n+1} - \sigma^{n}]/\Delta T$} \\
\texttt{(iv) 1st Correct: \\
 $~$\hskip40pt ${}^2\tilde S^{n+1} = \tilde S^n + \Delta T[0.4\dot{\tilde S}(\sigma^n,\dot\sigma^{n-1/2}) $\\
 $~$\hskip115pt     $+ 0.6\dot{\tilde S}({}^1\sigma^{n+1},{}^1\dot\sigma^{n+1/2})]$ } \\
\texttt{(iii) compute ${}^2u^{n+1}$ from ${}^2\tilde S^{n+1}$ \\
$~$\hskip40pt  ${}^2\dot u^{n+1/2} = [{}^2u^{n+1} - u^{n}]/\Delta T$, \\
$~$\hskip40pt  ${}^2\sigma^{n+1} = \sigma({}^2u^{n+1},{}^2\dot u^{n+1/2})$,\\
$~$\hskip40pt  ${}^2\dot \sigma^{n+1/2} = [{}^2\sigma^{n+1} - \sigma^{n}]/\Delta T$} \\
\texttt{(iv) 2nd Correct: \\
$~$\hskip40pt  $\tilde S^{n+1} = \tilde S^n + \Delta T[0.4\dot{\tilde S}(\sigma^n,\dot\sigma^{n-1/2})$ \\  $~$\hskip112pt  $+ 0.6\dot{\tilde S}({}^2\sigma^{n+1},{}^2\dot\sigma^{n+1/2})]$ }

By differencing the equations in this way, the
dominant nonviscous terms in the evolution
equations are accurate to second order (except for small effects due
to the use of the coefficients 0.4 and 0.6 in the corrector steps),
but small viscous terms involving time derivatives
are only accurate to first order in $\Delta T$.  Numerical
convergence tests show that our code is nearly second-order convergent
in space and time.  We find this to be sufficient for our purposes. 
When computing $\sigma_{\mu\nu}$, we first use Eq.~(\ref{sigma_def}) to get
the spatial components $\sigma_{ij}$.  The remaining components
$\sigma_{0\mu}$ are then obtained from the conditions
$u^{\mu}\sigma_{\mu\nu} = 0$.

As in most other Eulerian hydrodynamic codes, high velocities can easily
develop in the low-density regions near the surfaces of our stars.  The
method for evolving such regions in the absence of viscosity is described
in Paper~I.  Since calculation of the shear tensor involves taking
derivatives of the velocity field, we are unable to calculate it accurately
in the very low-density regions.  To ensure stability, we set $\eta = 0$ in
regions where $\rho_0 < 10^{-3} \rho_{0,\max}$.  Since these
low-density regions contain an insignificant amount of rest mass, this
prescription should not affect our evolutions.  We confirm this by
varying the cutoff density in several test problems and checking that 
the effect is negligible.

\subsection{Diagnostics}
\label{diagnostics}
Our most important diagnostics are the total mass-energy $M$
and angular momentum $J$. 
These are both defined by surface integrals at infinity~\cite{MandJ}:
\begin{eqnarray}
\label{Mdef}
M &=& {1\over 16\pi}\int_{S_{\infty}}\sqrt{\gamma}\gamma^{im}\gamma^{jn}
                                 (\gamma_{mn,j} - \gamma_{jn,m}) d^2S_i \\
\label{Jdef}
 J_i &=& {{1}\over{8\pi}}\varepsilon_{ij}{}^k \int_{S_{\infty}}
                         x^j K_k^m d^2S_m.
\end{eqnarray}
Using Gauss' Law, these surface integrals can be converted to volume
integrals:
\begin{eqnarray}
\label{Mfin}
M   &=& \int_V \Bigl(e^{5\phi}(\rho + {1\over 16\pi}\tilde A_{ij}\tilde A^{ij}
                             - {1\over 24\pi}K^2) \\
    & &  \quad - {1\over 16\pi}\tilde\Gamma^{ijk}\tilde\Gamma_{jik}
			     + {1-e^{\phi}\over 16\pi}\tilde R\Bigr) d^3x
			     \nonumber \\
J_i &=& \varepsilon_{ij}{}^k\int_V \Bigl({1\over 8\pi}\tilde A^j_k
        + x^j S_k \\
\nonumber
    & & \quad +  {1\over 12\pi}x^j K_{,k}
        - {1\over 16\pi}x^j\tilde\gamma^{lm}{}_{,k}\tilde A_{lm}\Bigr)
        e^{6\phi} d^3x\ .
\end{eqnarray}
In axisymmetry, the angular momentum integral simplifies to~\cite{Wald}
\begin{equation}
J_z = \int_V x S_y d^3x\ . \label{Jaxi}
\end{equation}
Henceforth, we drop the subscript $z$ since all angular momentum is
in the $z$-direction.

The mass $M$ and angular momentum $J$ in our grid should be strictly conserved
only in the absence of radiation (although gravitational radiation carries no
angular momentum in axisymmetry).  Since the energy and angular momentum
emitted in gravitational waves are negligible in our runs, this means that
$M$ and $J$ should be conserved in the no-cooling
limit.  They thus serve as useful code checks in this limit.  In
the rapid-cooling limit, the mass computed by the volume integral
(\ref{Mfin}) over the numerical grid will not be conserved---thermal
energy is carried off of the grid by thermal radiation.  The expected rate of
mass-energy decrease due to thermal energy loss can be computed by
differentiating Eq.~(\ref{Mfin}) with respect to time.  To lowest
order, we can ignore the effects of the quasi-stationary loss of
thermal energy on the
spacetime, and so ignore the time derivatives of field variables. 
Then only the first term in Eq.~(\ref{Mfin}) will be important.
\begin{eqnarray}
  \left.{dM\over dt}\right|_{\rm cooling}
  &=& {d\over dt}\int_V d^3x e^{5\phi}\rho + \cdots \nonumber \\
  &\approx& \int_V d^3x e^{5\phi}
  \left.{\partial\rho\over\partial t}\right|_{\rm cooling}\ ,
\end{eqnarray}
where $\partial\rho/\partial t|_{\rm cooling}$ is the component of
the time derivative
of $\rho$ caused by loss of internal energy due to cooling.  This quantity may
be computed by applying the chain rule to Eq.~(\ref{rho_source}):
\begin{equation}
\label{chain}
  \left.{dM\over dt}\right|_{\rm cooling} = \int_V
  \left.{\partial\rho\over\partial\epsilon}\right|_{\rho_0,u^0}
  \left.{\partial\epsilon\over\partial e_{\star}}\right|_{\rho_0,u^0}
  \left.{\partial e_{\star}\over\partial t}\right|_{\rm cooling}
  e^{5\phi} d^3x\ .
\end{equation}
Changes in $u^0$ are ignored because they represent a higher-order
influence on $dM/dt$.  The rate of change in $e_{\star}$ due to cooling
is given by the effective balance of heating and cooling
%($\Lambda = \Gamma_{\rm heat}$)
that characterizes the rapid-cooling limit (see Appendix A).  Thus,
$\partial_te_{\star}|_{\rm cooling}$
is minus the value of $\partial_te_{\star}$ due to viscous heating, i.e.
\begin{equation}
\label{e_cool}
\left.{\partial e_{\star}\over\partial t}\right|_{\rm cooling} = 
 	  -{2\over\Gamma}\alpha e^{6\phi}\eta
	  (\rho_0\epsilon)^{(1-\Gamma)/\Gamma}
	  \sigma^{\alpha\beta}\sigma_{\alpha\beta}
\end{equation}
[see Eq.~(\ref{evolve_estar})]. 
From equations (\ref{chain}) and (\ref{e_cool}), it is straightforward
to construct
\begin{eqnarray}
  \left.{dM\over dt}\right|_{\rm cooling} &=& -\int_V d^3x\ 2 e^{5\phi}
  \alpha e^{6\phi}\eta \sigma^{\alpha\beta}\sigma_{\alpha\beta} \\
  \nonumber
  & & \qquad \times \left({\rho_0\over\rho_{\star}}\right)
  \left[\Gamma e^{-12\phi}\left({\rho_{\star}\over \rho_0}\right)^2
    - \Gamma + 1\right]\ .
\end{eqnarray}
Finally, the quantity which is nearly conserved in the
rapid-cooling limit (up to losses due to gravitational radiation) is
\begin{equation}
\label{mtot}
  M_{\rm tot} = M + M_{\rm cooling} \equiv M - \int_0^t dt^{\prime} 
\left.{dM\over dt^{\prime}}\right|_{\rm cooling} \ .
\end{equation}

In both the no-cooling and rapid-cooling runs, 
we can divide $M$ into its constituent pieces: 
the rest mass $M_0$, internal energy mass $M_i$, kinetic energy $T$,
and gravitational potential energy $W$, defined by~\cite{cst92}
\begin{eqnarray}
\label{def:m0}
%M_0 &=& \int \rho_\star d^3x \\
%M_i &=& \int (\rho_0\epsilon) u^t\alpha e^{6\phi} d^3x \\
%T   &=& \int {1\over 2}\Omega T^t_{\varphi}\alpha e^{6\phi} d^3x \\
%W   &=& M - M_0 - M_i - T \label{def:W}
M_0 &=& \int_V \rho_0 d{\cal V} \label{def:M_0} \\
M_i &=& \int_V (\rho_0\epsilon) d{\cal V} \\
T   &=& \int_V {1\over 2}\Omega T^0_{\varphi} (u^0)^{-1} d{\cal V} \\
W   &=& M - M_0 - M_i - T\ , \label{def:W}
\end{eqnarray}
where $d{\cal V} = \alpha u^0 e^{6\phi} d^3x$ is the proper 3-volume
element. 
To study the effects of heating, it is useful to break up the internal
energy $\epsilon$ into its ``cold'' component
$\epsilon_0 = \rho_0^{\Gamma-1}/(\Gamma - 1)$, and its ``thermal''
component $\epsilon_{\rm heat} = \epsilon - \epsilon_0$.  Then we
can break up $M_i$ into cold and hot components
\begin{eqnarray}
M_{ic} &=& \int_V (\rho_0\epsilon_0) d{\cal V} \\
M_{ih} &=& \int_V (\rho_0\epsilon_{\rm heat}) d{\cal V}\ .
\label{mih}
\end{eqnarray}
Note that in the rapid-cooling limit, $M_{ih} = 0$.

Finally, we also compute the circulation along
closed curves.  For a closed curve $c$ with tangent
vector $\lambda^{\mu}$, the circulation is defined to be
\begin{equation}
  {\cal C}(c) = \oint_{c} h u_{\mu} \lambda^{\mu} d\zeta\ , 
\label{circ_define}
\end{equation}
where $\zeta$ parameterizes points on $c$
[i.e. $\lambda^{\mu} = (\partial/\partial\zeta)^{\mu}$]. According to
the Kelvin-Helmholtz theorem, the
circulation ${\cal C}$
will be conserved in the absence of viscosity if $c$
moves with the fluid and if the fluid is barotropic [$P = P(\rho_0)$]. 
When viscosity is present or the equation of state is more general,
as in the case of nonisentropic flow, the Navier-Stokes equations give
\begin{equation}
  {d{\cal C}\over d\tau} = -\oint_{c} \lambda^{\mu}\rho_0^{-1}
  \left[P_{,\mu} - 2\left(\eta\sigma_{\mu}{}^{\nu}\right)_{;\nu}\right]
  d\zeta\ .
  \label{evolve_circ}
\end{equation}
If $\eta = 0$, the second term in the integrand vanishes,
and if $P = P(\rho_0)$, the remaining term is an exact
differential.  Then $d{\cal C}/d\tau$ integrates to zero,
in accord with the Kelvin-Helmoltz theorem.

In axisymmetry, we choose to evaluate ${\cal C}$ on circular rings
on the equatorial $z = 0$ plane, so that $\zeta = \varphi$.  The ring
$c$ intersects our 2D grid at a point on the $x$-axis. 
By our symmetries, the curve $c$
will always remain circular and always remain at $z = 0$, so the
Lagrangian point representing $c$ only moves in $x$.
%Also, axisymmetry ensures
%that each point on $c$ moves in the same way, so that hypersurfaces
%of constant $t$ are also hypersurfaces of constant $\tau$. 
Evolution in $t$ and in $\tau$ are simply related by $d/d\tau = u^0 d/dt$. 
Since the system is axisymmetric, the integrand, being a scalar, is
constant along $c$, so Eqs.~(\ref{circ_define}) and
(\ref{evolve_circ}) simplify to
\begin{eqnarray}
  {\cal C} &=& 2\pi h u_{\varphi} = 2\pi h x u_y \ \ \ \ \mbox{(axisym)} \\ 
  {d{\cal C}\over dt} &=& {1\over u^0}{d{\cal C}\over d\tau} = 
  {4\pi\over\rho_0u^0}
  (\eta\sigma_{\varphi}{}^{\nu})_{;\nu} \nonumber \\
  &=& {4\pi\over \rho_{\star}x}
  \left[(x^2\alpha e^{6\phi}\eta\sigma_y{}^0)_{,t} + (x^2\alpha e^{6\phi}\eta\sigma_y{}^A)_{,A}\right]
  \label{Circ_fin}\ .
\end{eqnarray}
Hence the quantity ${\cal C}_{\rm tot}$ given by
\begin{equation}
  {\cal C}_{\rm tot} = {\cal C} + {\cal C}_{\rm vis} \equiv {\cal C} 
  - \int_0^t dt^{\prime} {d{\cal C}\over dt^{\prime}} \label{ctot}
\end{equation}
is conserved, even in the presence of viscosity.

Finally, we compute the Hamiltonian and momentum constraint violations
[given by equations (16) and (17) of Paper~I].  We monitor the L2 norm
of the violation of each constraint.  We compute the L2 norm of a
gridfunction $g$ by summing over every gridpoint $i$:
\begin{equation}
  L2(g) = \sqrt{\sum_i g_i^2}\ .
  \label{L2}
\end{equation}
The constraint violations are normalized as described in Paper~I
[Eqs.~(59) and (60)].

\section{Code Tests}

\begin{table*}
\caption{Initial equilibrium models for code tests.}
\begin{center}
\tabcolsep 0.07in
\begin{tabular}{c c c c c c c c c}
\hline 
\hline
\vspace{0.03in}
Star & $M$ & $M_0$ & $R_{\rm eq}/M$ & $\rho_{0,{\rm max}}$${}^{\rm a}$ & 
$T/|W|$ & $\Omega_{\rm eq}/\Omega_c$${}^{\rm b}$ & 
$P_{\rm rot}/M$${}^{\rm c}$ & $\mathcal{R}_{\rm pe}$${}^{\rm d}$  \\
\hline
A & 0.170 & 0.186 & 4.10 & 0.241 & 0.032 & 1.0   & 155  & 0.88 \\
B & 0.171 & 0.187 & 3.48 & 0.363 & 0.031 & 1.0   & 125  & 0.87 \\
C & 0.183 & 0.200 & 4.53 & 0.155 & 0.095 & 0.346 & 60.6 & 0.73 \\
D & 0.241 & 0.260 & 5.47 & 0.061 & 0.234 & 0.383 & 52.4 & 0.37 \\
E & 0.259 & 0.277 & 5.92 & 0.045 & 0.263 & 0.381 & 57.4 & 0.28 \\
\hline
\hline
\end{tabular}
\end{center}
\vskip 6pt
\begin{minipage}{11cm}
\raggedright
${}^{\rm a}$ {Maximum rest-mass density.  This does not correspond to the
center of the star for hypermassive, toroidal models D and E.}
\\
${}^{\rm b}$ {Ratio of $\Omega$ at the equatorial surface to $\Omega$ at 
the center.}
\\
${}^{\rm c}$ {Initial central rotation period.}
\\
${}^{\rm d}$ {Ratio of polar to equatorial coordinate radius.}
\end{minipage}
\end{table*} 

%These are the TOV stars that we're really not discussing
%A & 0.157 & 0.171 & 0.200 & 0.0   & -     & -     & 1.00 \\
%B & 0.162 & 0.178 & 0.400 & 0.0   & -     & -     & 1.00 \\

In Paper~I, we presented our relativistic hydrodynamic code.  This
code evolves the coupled Einstein field relativistic hydrodynamic 
system on 3D grids,
assuming perfect-fluid hydrodynamics.  We demonstrated the ability
of our code to distinguish stable from unstable relativistic polytropes, to
accurately follow gravitational collapse of rotating stars, 
and to accurately evolve
binary polytropes in quasi-stationary circular orbits.
When black holes appear on our grid, we can employ excision to 
remove the spacetime singularities from our grid. 
Tests of our black hole excision algorithm using this code 
were reported in \cite{ybs01} for single rotating black holes in vacuum
spacetimes and in \cite{dsy03} for black holes that arise during the collapse
of hydrodynamic matter.
In this section, we test the adaptations of this code which 
force axisymmetric evolution and the modifications which allow a physical 
viscosity.  Simulations performed with our axisymmetric code 
show that stable and unstable TOV stars are correctly distinguished.
The code also achieves approximate second order convergence in the
evolution of
linear gravitational waves and TOV stars.  Below, we describe test 
runs on rotating stars in some detail.  First, we consider stable and 
unstable uniformly rotating stars, as well as a stable differentially 
rotating star, in axisymmetry and without viscosity.  We then test 
the sensitivity of our code to nonaxisymmetric {\it dynamical} bar formation.  Finally,
we check that physical viscosity is implemented correctly by
considering stable uniformly and differentially rotating models.  
A summary of the models used for these code tests is given in Table~I.
The models are initially $n=1$ equilibrium polytropes and are evolved 
using a $\Gamma$-law equation of state with $\Gamma =2$ 
[see Eq.~(\ref{ideal_P})]. 
Initial data for all of these models  
are obtained from the relativistic, rotating star equilibrium code
of~\cite{cst92}.  Stars A and B were also 
studied in Paper I \cite{note:modelsAB}. For the differentially rotating 
stars C, D, and E, we choose 
the initial rotation profile
\begin{equation}
  u^t u_{\varphi} = R_{\rm eq}^2 A^2(\Omega_c - \Omega)\ ,
  \label{rot_law}
\end{equation}
where $\Omega$ is the angular velocity of the fluid, $\Omega_c$ is the
value of $\Omega$ at the center and all along the rotation axis, $R_{\rm eq}$
is the equatorial
coordinate radius, and the parameter $A$, which measures the degree of
differential rotation, is chosen to be unity. 
In the Newtonian limit, Eq.~(\ref{rot_law}) reduces to the so-called
``$j$-constant'' law~\cite{eriguchi85}
\begin{equation}
  \Omega = {\Omega_c\over 1 + {\varpi^2\over R_{\rm eq}^2 A^2}}\ .
\end{equation}
We note that the maximum mass, $n = 1$ TOV polytrope has mass
$M = 0.164$ and compaction $R_{\rm eq}/M = 3.59$~\cite{cst92}. 
All of the axisymmetric tests in this section were performed with a
modest resolution of $64 \times 64$ and outer boundaries at about $12 M$.   
Passing these tests successfully with modest resolution helps establish the
robustness of our code.

\subsection{Tests in axisymmetry}
\label{axitest}    
\begin{figure}
\includegraphics[width=8cm]{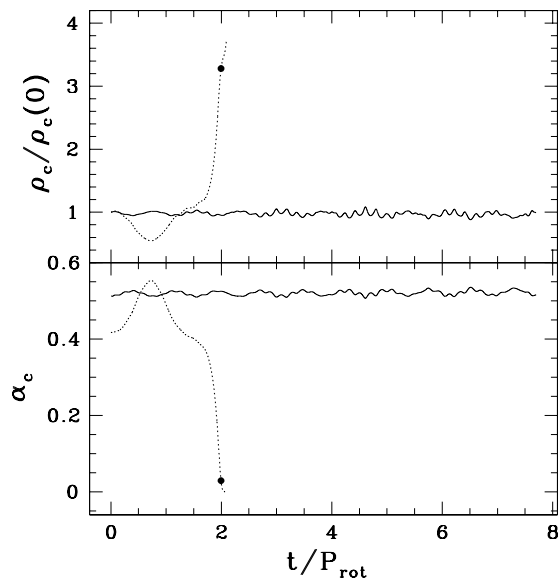}
\vskip 0.5cm
\caption{Axisymmetric evolution of uniformly rotating stars. Star~A (solid 
lines) is stable, while star~B (dotted lines) is unstable to collapse. 
The upper window shows
the central density normalized to its initial value, while the lower
gives the central lapse.  The solid dot indicates the first appearance of
an apparent horizon during the collapse of star~B.}
\label{fig:unif}
\end{figure}

%\subsubsection{Uniform Rotation}
To demonstrate that our ``axisymmetrized'' code can distinguish stable and 
unstable uniformly rotating models, we consider stars A and B. 
These stars lie along a sequence of constant angular momentum, uniformly 
rotating stars.  As described in Paper~I, star~A lies to the left of 
the turning point on the $M$-$\rho_c$ equilibrium curve, while star~B lies to the right.
Then by the theorem of Friedman, Ipser, and Sorkin \cite{fis88}, star~B is 
secularly unstable to radial perturbations, while star~A is stable.  
Since the onset of dynamical radial instability is very close 
to the onset of secular instability for such sequences \cite{sbs00}, we 
expect that star~A will be stable to collapse, while star~B will be 
dynamically unstable.  
When evolved in axisymmetry, star~A persists for more than $7 P_{\rm rot}$ 
without significant changes in structure, where the central rotation 
period is  
\begin{equation}
\label{Prot}
P_{\rm rot} \equiv \frac{2\pi}{\Omega_c(t=0)} \ .
\end{equation}
The oscillations in $\rho_c$, which 
correspond to radial pulsations, have an amplitude of $\lesssim\! 7\%$.  For 
this run, the Hamiltonian and momentum
constraints are satisfied to within $2\%$, while $M$
is conserved to better than $1\%$. 
%constraints were satisfied to within $\sim\! 2\%$, while $M$, $M_0$, and $J$ 
%were conserved to better than $1\%$. 
Meanwhile, the unstable uniformly rotating star (star B) collapses, with an 
apparent horizon first appearing at time $t \simeq 2 P_{\rm rot}$, 
corresponding to 21.3 light crossing times of the grid. 
At this time, the constraints are satisfied to within 
$6\%$ and $M$ is conserved to within $3\%$. Thus, stable and
unstable uniformly rotating stars are clearly distinguished even at this
moderate resolution.  Figure~\ref{fig:unif} summarizes the results for these 
two runs.

\vskip 1cm

%\subsubsection{Differential rotation.}  
Next we consider the evolution of a differentially rotating star using
our axisymmetric code.  We evolve star~C for a time $\ga 15 P_{\rm rot}$.  
Throughout the simulation, all constraints are satisfied to better than $4.5\%$, 
while $M$, $J$, and $M_0$ are conserved to within $3.5\%$ ($J$ and $M_0$ 
decrease due to flow beyond the outer boundaries).  In the absence of a 
dissipative mechanism 
to brake the differential rotation, the structure of the equilibrium 
star should not 
change.  We find that we can numerically hold this equilibrium state for
$15 P_{\rm rot}$.  Note that the small amount of Kreiss-Oliger dissipation
employed for numerical stability does not alter the rotation profile of the star.
%After this time, $\Omega$ grows unacceptably large 
%for the inner few gridpoints and $\rho_0$ begins to vary strongly near the 
%axis on the scale of the gridspacing.  These problems worsen over the course 
%of the run.  Evidently, Kreiss-Oliger dissipation is not enough to prevent 
%the problems which arise near the axis for very long runs with our code.
After this time, inaccuracies at the center, manifested by growth in
$\Omega$ and high-frequency oscillations in $\rho_0$, begin to grow.  
We monitor the evolution of the 
circulation for three different fluid elements chosen at the following 
initial locations in the equatorial plane: (a) $r=R_{\rm eq}/4$, 
(b) $r=R_{\rm eq}/2$, and (c) $r=3R_{\rm eq}/4$, where $R_{\rm eq}$ is 
the initial radius of the star. We find that, for 
$t \lesssim 22.5 P_{\rm rot}$, the circulation is conserved to better than 
$5\%$ for all three of these points.  After this time, the same
inaccuracies cause the circulation to deviate from its initial value.
%Eventually, however, these circulation 
%values deviate by as much as $30\%$ from their initial values.  This indicates 
%that the azimuthal velocity field becomes inaccurate for sufficiently long 
%runs (although the total angular momentum is approximately conserved).

Because viscosity tends to smooth irregularities in the velocity
field, the problems near the axis and the inaccuracy of the azimuthal 
velocity can be controlled by a small shear viscosity.  
To test this we evolve star~C for 
$\sim 60 P_{\rm rot}$ with a very small shear viscosity (such 
that $\tau_{\rm vis} \simeq 550 P_{\rm rot}$).  Because $\tau_{\rm vis}$ is
so much greater than the length of the simulation, the small viscosity does not 
significantly alter
the structure of the star. We find that the behavior of 
$\Omega$ improves considerably, while the small-scale variations in $\rho_0$ 
near the 
axis do not occur.  In addition, the circulation values for the same three 
points which we studied in the previous case are conserved to within $5\%$ for 
more than $50 P_{\rm rot}$.  Thus, even a tiny shear viscosity  
significantly lengthens the period during which our runs are accurate.  As 
we will 
describe below, the presence of a small shear viscosity allows us to 
evolve axisymmetric models accurately for hundreds of $P_{\rm rot}$, 
corresponding to thousands of $M$. 
       
%\includegraphics[width=8cm]{denprof.ps}
%Caption: Equatorial density profiles taken at selected times normalized to 
%the initial central density for a run of star C in axisymmetry with 
%$\nu_{\rm P} = 0$.  The distribution of mass within the star changes 
%little over many rotation periods.

\subsection{Tests of dynamical bar mode sensitivity} 
We now demonstrate that our 3D code is sensitive to the 
nonaxisymmetric {\em dynamical} 
bar-mode instability. This sensitivity is
important because the results of axisymmetric runs for a particular
case will only be valid physically if it can be demonstrated that nonaxisymmetric
modes do not develop in the corresponding 3D evolution.  We consider two 
models, D and E, which we expect to be dynamically stable and unstable to bars,
respectively.  This expectation is based on earlier 3+1 fully relativistic 
evolutions of these stars by Shibata, Baumgarte, and Shapiro \cite{sbs02}, 
who studied the formation of bars.
Both models are hypermassive, toroidal configurations
with high values of $T/|W|$ (0.230 for D and 0.258 for E).  
These models are identical to models D1 and D2 considered 
in \cite{sbs02}.  To test for bars, 
we add a nonaxisymmetric density perturbation
of the following form to the axisymmetric initial data:
\begin{equation}
\label{seedbar}
\rho = \rho_0\left(1 + \delta_b\frac{x^2 - y^2}{R_{\rm eq}^2}\right),
\end{equation}
where $\delta_b$ parameterizes the strength of the initial bar deformation.
We choose $\delta_b=0.1$ for both models D and E.
We then re-solve the constraint equations as in \cite{Shibata:1999aa} 
to ensure that they are satisfied on the initial time slice.  
The growth of a bar is indicated by the quadrupole 
diagnostic (see \cite{sbs03}),
\begin{equation}
Q = \langle e^{im\varphi} \rangle_{m=2} = \frac{1}{M_0} \int d^3x \rho_*
\frac{(x^2 - y^2) + 2ixy}{x^2 + y^2}.
\end{equation}  
We will take $|Q| = \sqrt{Q^*Q}$ as a measure of the magnitude of the 
bar deformation. 

\begin{figure}
\includegraphics[width=8cm]{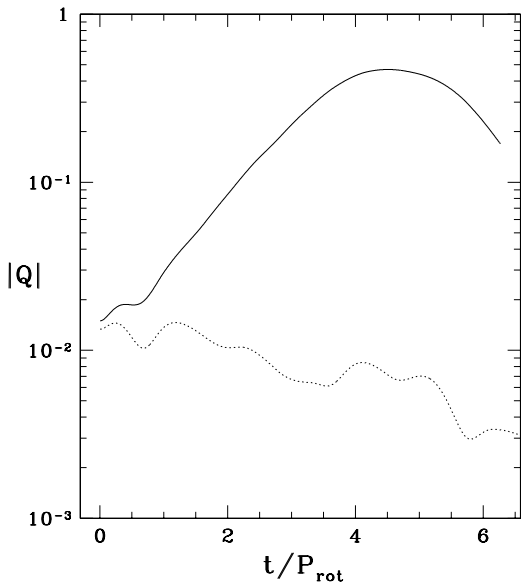}
\vskip 0.5cm
\caption{Quadrupole diagnostic for evolutions of rapidly rotating 
hypermassive stars. Star~D is stable and star~E is unstable
to dynamical bar formation.  The initial $m=2$ perturbation decays
for star D (dotted line) but grows for star E (solid line).  Note that, 
for each curve, the time 
axis is normalized by $P_{\rm rot}$ which differs for the two stars.  }
\label{fig:q}
\end{figure}

We evolve star~D for a time $8.9 P_{\rm rot}$, during which 
$M$ and $J$ were conserved to within $0.7\%$ and all constraints were 
satisfied to within $2\%$.  For Star E, the run was terminated after
$6.3 P_{\rm rot}$ and $M$ and $J$ were conserved to within 
$1.0\%$, while
constraint violations were $\lesssim\! 5.5 \%$.  Both runs were performed 
in $\pi$-symmetry on uniform grids with resolution 
$128 \times 64 \times32$.  The outer boundaries in the $x$-$y$ plane
were at $16.6 M$ for star D and $19.3 M$ for star E. The 
results are shown in Fig.~\ref{fig:q}.  This test clearly shows the growth of the 
bar mode for star E, while star D does not form a bar even with 
the substantial initial perturbation.  

%Run parameters for D:
%Resolution $128 \times 64 \times32$, with $\nu = 0$ and boundaries 
%$x \in (-4,4)$, $y \in (0,4)$, and $z \in (0,2)$. Run parameters for E:
%Resolution $128 \times 64 \times32$, with $\nu = 0$ and boundaries 
%$x \in (-5,5)$, $y \in (0,5)$, and $z \in (0,2.5)$.

\vskip 1cm

\subsection{Tests with viscosity}
\label{visctest}
\begin{figure}
\includegraphics[width=8cm]{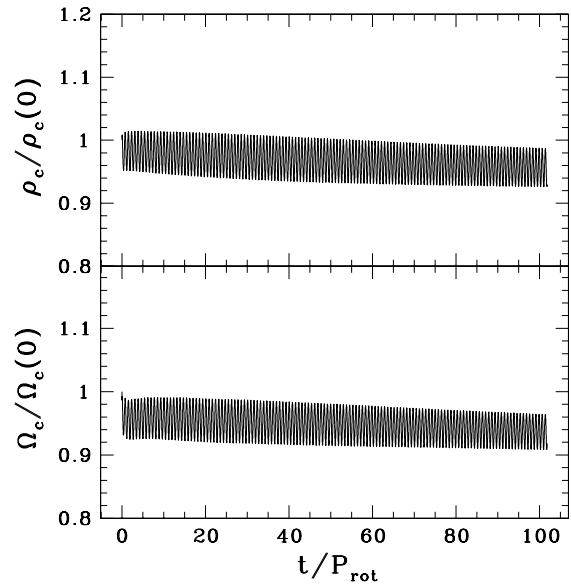}
\vskip 0.5cm
\caption{Evolution of the stable, uniformly rotating star~A with high 
viscosity.
The central density (shown in the upper window) and central angular
velocity (lower window) oscillate without changing appreciably for
over 100 rotation periods ($15,500 M$).}
\label{fig:visunif}
\end{figure}

%\subsubsection{Uniform Rotation} 
As a first test of our shear viscosity implementation, we demonstrate
that uniformly rotating configurations are unaffected by the presence
of even a large viscosity.  We evolve the uniformly rotating, stable 
star~A with 
$\nu_{\rm P} = 0.2$ (such that $\tau_{\rm vis} \simeq 0.09 P_{\rm rot}$) 
for $\sim\!\! 100 P_{\rm rot} \sim 15,500 M$.  The mass $M$ is 
conserved to within $0.1\%$, and $J$ to within 
$1.5\%$.  (Note that, even in axisymmetry, $J$ is not identically 
conserved by our finite differencing scheme when viscosity is present.)
All constraints are satisfied to better than $1.1\%$ for the duration of 
this run.  The resulting evolution of the central rest-mass density and 
central angular velocity are shown in Fig.~\ref{fig:visunif}.  These 
quantities 
oscillate on the radial oscillation timescale ($\sim\!\tau_{\rm FF}$) with 
amplitudes of several percent, and this run encompassed roughly 120 
oscillation periods.  Because the oscillations are radial, they are not 
appreciably damped by the shear viscosity.  For the entire run, the 
average values of $\rho_c$ and 
$\Omega_c$ drop by about $4.5\%$ and $6.5\%$, respectively.  
These small deviations are due to the accumulated numerical 
error of the finite differencing and are reduced by increasing
resolution.
Since the star does not change appreciably over many viscous timescales, 
our code simulates the correct physical behavior for this case.  

\vskip 1cm
%Run parameters for this case: Resolution $64 \times 64$,
%with $\nu = 0.2$ and boundaries $x,z \in (0,2)$.

\begin{figure}
\includegraphics[width=8cm]{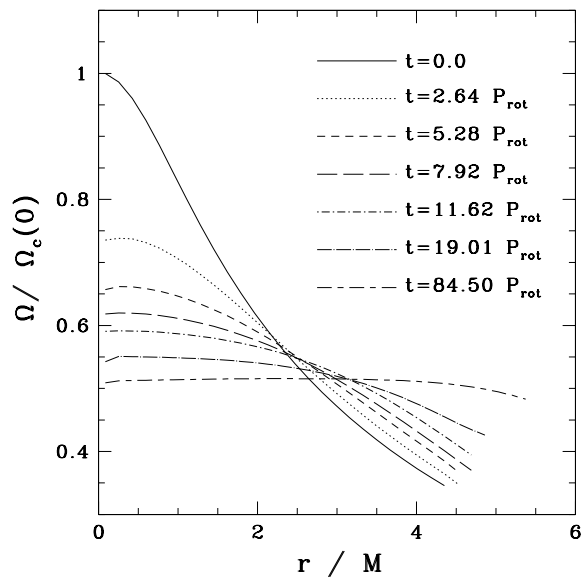}
\vskip 0.5cm
\caption{Angular velocity profiles in the equatorial frame at selected times 
for star C with $\tau_{\rm vis} \approx 5.5 P_{\rm rot}$.  The presence of 
viscosity drives the star to uniform rotation on a viscous timescale.}
\label{fig:omprof}
\end{figure}

\begin{figure}
\includegraphics[width=8cm]{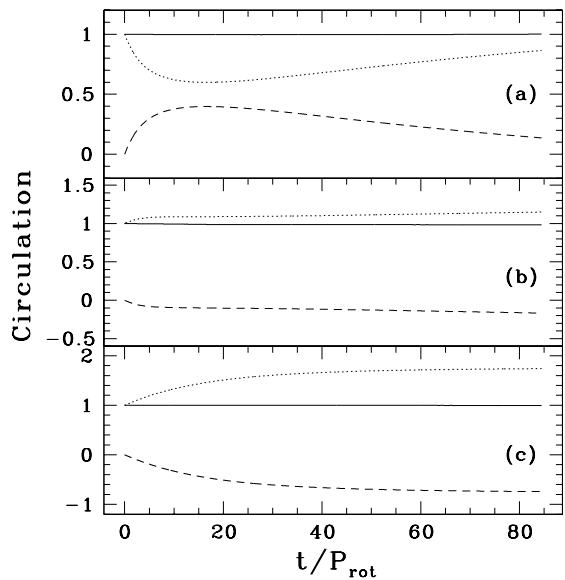}
\vskip 0.5cm
\caption{Evolution of the circulation for three selected fluid elements of 
star C in axisymmetry (using 
$\nu_{\rm P} = 0.015$).  The dotted line gives the  
circulation ${\cal C}$, the dashed line gives the time integrated contribution due
to viscosity ${\cal C}_{\rm vis}$, and the solid line gives their sum ${\cal C}$, 
which is well-conserved [see Eq.(\ref{ctot})].
Each quantity is normalized by the corresponding initial circulation.}
\label{fig:circall}
\end{figure}

%\subsubsection{Differential Rotation}
We now test the viscous evolution of the differentially rotating star~C.
We choose viscosity
$\nu_{\rm P} = 0.015$, so that Eq.~(\ref{tvis_P}) gives 
$\tau_{\rm vis} \approx 5.5 P_{\rm rot}$. In axisymmetry, we ran this case 
for $84.5 P_{\rm rot} = 15.4 \tau_{\rm vis} = 5,120 M$, during which time 
$M$ and $J$ are conserved to within $0.4\%$ 
while all of the constraints are satisfied to better than $1.1\%$.  
Figure~\ref{fig:omprof} 
shows several snapshots of the angular velocity profile in the equatorial 
plane for the 2D case taken 
at various times.  This clearly shows that the presence of viscosity drives 
the star toward uniform rotation.  As a quantitative test of the
action of viscosity, we check that the circulation evolves according 
to Eq.~(\ref{Circ_fin}).  Choosing three fluid elements in the initial 
configuration, we track these fluid elements and calculate the circulation
$\cal{C}$ for each one, as well as the time-integrated contributions from 
viscosity ${\cal C}_{\rm vis}$ [see Eq.~(\ref{ctot})].  The fluid elements 
are chosen at the same locations as for 
the inviscid test of star C in Section \ref{axitest}.  The results are shown 
in Fig.~\ref{fig:circall}, which gives the circulation, the viscous contribution, 
and their sum, ${\cal C}_{\rm tot}$.  For all three cases, ${\cal C}_{\rm tot}$  
is conserved to better than $2\%$ 
for the entire run. Thus, angular momentum is transported correctly for many 
tens of rotation periods (thousands of $M$) when a significant 
shear viscosity is present.

\vskip 1cm

\vskip 1cm
    
\begin{figure}
\includegraphics[width=8cm]{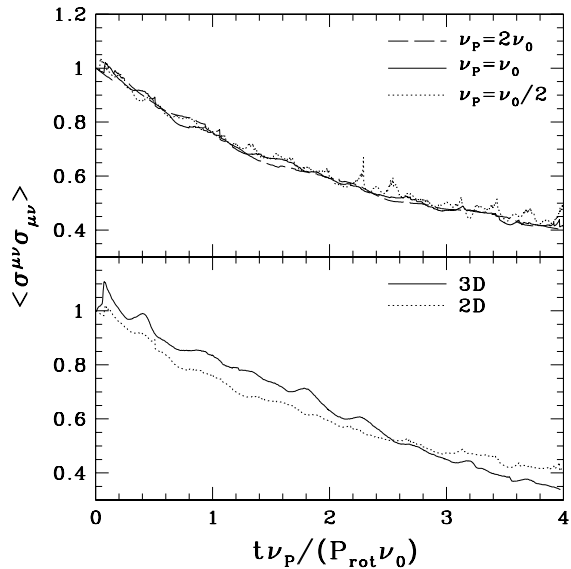}
\vskip 0.5cm
\caption{Energy dissipation rate $\langle\sigma^{\mu\nu}\sigma_{\mu\nu}\rangle$,
normalized to its initial value, for several runs with star C.  
The upper window demonstrates the proper scaling of our solutions with 
$\nu_{\rm P}$. Note that the time axis is scaled according to the appropriate value of 
$\nu_{\rm P}$.  The lower window compares 
$\langle\sigma^{\mu\nu}\sigma_{\mu\nu}\rangle$ for runs in axisymmetry (2D) and 
$\pi$-symmetry (3D), both with $\nu_{\rm P} = 0.015 = \nu_0$.  These evolutions agree 
fairly well.}
\label{fig:sigsig}
\end{figure}

We also used this case to test the scaling behavior of our solutions with 
$\nu_{\rm P}$.  Results are shown in the upper window of 
Fig.~\ref{fig:sigsig}, which gives the evolution
of $\langle\sigma^{\mu\nu}\sigma_{\mu\nu}\rangle$ for several values of 
$\nu_{\rm P}$ versus scaled time.  We define the energy dissipation rate 
via shear viscosity, 
$\langle\sigma^{\mu\nu}\sigma_{\mu\nu}\rangle$,
as in Section~\ref{justification} [see Eq.~(\ref{evolve_estar})].  This 
quantity decays due to the action of viscosity.  The figure shows that our
solutions obey the proper scaling with $\nu_{\rm P}$, i.e. they evolve
identically but on a timescale inversely proportional to the adopted
viscosity ($\nu_P$).  Hence our results can all be scaled to the much smaller 
viscosities likely to be appropriate for physically realistic viscosity in stars.
We provide further demonstration of scaling in Section \ref{2Dcooling} (see 
Fig.~\ref{fig:scaling}).

\vskip 1cm
   
The results of the axisymmetric run with $\nu_{\rm P} = \nu_0$ agree fairly well
with a 3D, $\pi$-symmetric run performed for the same model.  This 3D run employed
$64\times 32\times 32$ grid zones, giving only half of the resolution of the 
axisymmetric run.  We terminated 
the 3D run after $6.35 P_{\rm rot}$.  $M$ is conserved to within $0.4\%$, 
while the constraint violations are $\lesssim\! 3.6 \%$.  Because of the lower 
resolution in this case, $3.2\%$ of the total angular momentum is lost
(as opposed to $0.4\%$ for the much longer axisymmetric run).  A comparison of 
$\langle\sigma^{\mu\nu}\sigma_{\mu\nu}\rangle$ for the 2D and 3D cases is 
plotted in the lower window of Fig.~\ref{fig:sigsig}, and shows good agreement.  

\section{Dynamical Evolutions}

\subsection{Introduction and discussion of models}

\begin{table*}
\caption{Initial Models}
\begin{center}
\begin{tabular}{c c c c c c c c c c}
\hline 
\hline
\vspace{0.03in}
Case & \small{$M_0/M_{0,\rm{TOV}}$${}^{\rm a}$} & 
\small{$M_0/M_{0,\rm{sup}}$${}^{\rm b}$}  & 
$M$ & $R_{\rm eq}/M$ & $J/M^2$ & $T/|W|$ & 
$\Omega_{\rm eq}/\Omega_c$ &
$P_{\rm rot}/M$ & $\nu_P$${}^{\rm c}$  \\
\hline\hline
  I  & 1.69 & 1.38 & 0.279 & 4.48  &  1.0  & 0.249 & 0.33 & 38.4 & 0.2 \\
\hline
  II   & 1.39 & 1.13 & 0.228 & 4.40 & 0.85 & 0.188 & 0.32 & 41.3 & 0.07 \\ 
\hline
  III  & 1.39 & 1.13 & 0.232 & 5.54 & 1.0 & 0.224 & 0.37 & 54.2 & 0.15 \\
\hline
  IV   & 1.39 & 1.13 & 0.234 & 6.27 & 1.1 & 0.244 & 0.31 & 63.3 & 0.2 \\
\hline
  V    & 1.0  & 0.81 & 0.168 & 8.12 & 1.0 & 0.181 & 0.40 & 103 & 0.15 \\
\hline \hline
\end{tabular}
\end{center}
\vskip 12pt
\begin{minipage}{16cm}
\raggedright
${}^{\rm a}$ {If this ratio is greater than unity, the star's mass exceeds the 
TOV limit for $n=1$ polytropes ($M_{0,\rm TOV}=0.180$).}
\\
${}^{\rm b}$ {If this ratio is greater than unity, the star's mass exceeds the 
uniformly rotating (supramassive) upper limit ($M_{0,\rm sup}=0.221$)
and is therefore hypermassive.}
\\
${}^{\rm c}$ {The values of $\nu_P$ are chosen such that the viscous timescale 
$\tau_{\rm vis}\approx 3 P_{\rm rot} \sim 10 \tau_{\rm FF}$.}
\end{minipage}
\label{tab:models}
\end{table*}

\begin{figure}
\vskip 1cm
\includegraphics[width=8cm]{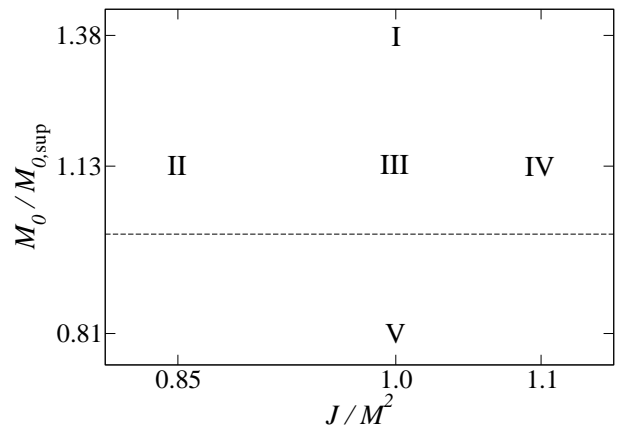}
\vskip 0.5cm
\caption{Rest mass $M_0$ and spin parameter $J/M^2$ for the five selected models
in Table~\ref{tab:models}. The dashed line denotes the
mass limit of uniformly rotating supramassive $n=1$ polytropes, $M_0=M_{0,{\rm sup}}$. 
All stars
above this line are hypermassive and require differential rotation
to be in hydrostatic equilibrium.}
\label{fig:models}
\end{figure}

Having shown simulations for several test models, we now present the 
evolution of five differentially rotating, dynamically stable stellar
models in which viscosity changes the 
structure of the stars in nontrivial ways. Our models are summarized in 
Table~\ref{tab:models} and Fig.~\ref{fig:models}. We first perform 
short, 3D simulations without viscosity on all the five models to make 
sure that they are all dynamically stable to bar formation.
Each of them is then evolved with our axisymmetric code in 
both the rapid and no-cooling limits described in Section~\ref{rad_cooling}. 
The initial data for the five stars are again computed with the 
relativistic equilibrium code of~\cite{cst92}.
The stars obey an $n=1$ polytropic equation of state $P=\rho_0^2$.
We adopt the rotation law given by Eq.~(\ref{rot_law}) with $A=1$.
This rotation law has been found to be 
a good approximation to the angular velocity profile of 
proto-neutron stars formed from core collapse~\cite{vpce03}. 
In the case of a binary neutron star merger, the remnant can form 
a dynamically stable hypermassive neutron star provided the remnant
mass does not exceed about 1.7 times the maximum 
mass of a nonrotating spherical star~\cite{Shibata:1999wm}. 
Our adopted rotation law is also found to be a reasonably good approximation 
to the angular velocity profile of these hypermassive neutron 
stars~\cite{lbs03}. 

In all of our axisymmetric calculations, we use a grid size $128\times 128$ 
with an outer boundary at $14M$ for the most massive and compact star (star~I), 
and $24M$ for the least massive and compact star (star~V).
Initially, the equatorial 
radii of the stars are only about $R_{\rm eq} \approx 5 M$. However, 
viscosity causes the outer layers to expand 
and, in some cases, we find that a few percent of rest mass is lost due to 
material flowing out of the grid. In each model, we choose the value of
the viscosity coefficient $\nu_P$ such that the viscous timescale defined 
by Eq.~(\ref{tvis_P}) is 
$\tau_{\rm vis}\approx 3 P_{\rm rot}\sim 10 \tau_{\rm FF}$. 
With this moderate strength of viscosity, we need to evolve the stars 
for 100--200$P_{\rm rot}$ in most cases to follow the complete secular 
evolution and determine the final fate of the stars. The reason is that 
in most cases, viscosity generates a low-density envelope around the 
central core. Since our viscosity law has $\eta \propto P$, the viscosity 
in the low-density region is small. 
(The density throughout the envelope is greater, however, than the cutoff density below
which $\eta = 0$.) Hence the effective viscous timescale increases with 
time and it takes longer for the stars to achieve a final state. 

\begin{table*}
\caption{Summary of simulations.}
\begin{center}
\begin{tabular}{c c c c c c c c c c c}
\hline
\hline
\vspace{0.03in}
Case & 2D/3D & \small{Cooling} & $t_{\rm final}/P_{\rm rot}{}^{\rm a}$ & 
Initial $T/|W|$ 
& Final $T/|W|$ & 
$\langle P/\rho^{\Gamma}\rangle$${}^{\rm b}$ & Fate & $J_h/M_h^2{\ }^{\rm c}$ & 
$M_{0,\rm disk}/M_0$ & $J_{\rm disk}/J$ \\
\hline\hline
  I  & 2D & no  & 28.9 & 0.25  & 0.09${}^{\rm d}$ & 3.9 & BH + disk & 0.6 
& 0.23 & 0.65 \\
     & 2D & yes & 13.7 &   & 0.15${}^{\rm d}$ & 1.0 & BH + disk & 0.8 & 
0.21 & 0.55 \\
     & 3D & yes & 9.3 &   & 0.21${}^{\rm d}$ & 1.0 & BH + disk & 0.8 & 
0.18 & 0.47 \\
\hline
  II & 2D & no & 286 & 0.19 & 0.09 & 2.8 & star + disk & -- & 0.15 & 0.56 \\
     & 2D & yes & 57.7 &   & 0.14${}^{\rm d}$ & 1.0  & BH + disk & 0.7 &
0.10 & 0.36 \\
\hline
 III & 2D & no & 105 & 0.22 & 0.09 & 4.3 & star + disk & -- &  0.21 & 0.68 \\
     & 2D & yes & 315 &  & 0.12 & 1.0 & star + disk & -- & 0.15 & 0.58  \\
\hline
  IV & 2D & no & 99 & 0.23 & 0.10 & 7.3 & star + disk & -- & 0.25 & 0.76 \\
     & 2D & yes & 235 &  & 0.13 & 1.0  & star + disk  & -- & 0.17 & 0.62 \\
     & 3D & yes & 11.5 &  & -- & 1.0  & no bar & --  & -- & -- \\
\hline
  V  & 2D & no & 105 & 0.18 & 0.09 & 3.7 & star + disk & -- & 0.13 & 0.52 \\
     & 2D & yes & 171  & & 0.13  & 1.0  & star + disk & -- & 0.09  & 0.38 \\
\hline
\hline
\end{tabular}
\end{center}
\vskip 12pt
\begin{minipage}{16cm}
\raggedright
${}^{\rm a}$ {The time at which the simulation was terminated.} \\
${}^{\rm b}$ {This quantity corresponds to an average of $P/\rho^{\Gamma}$ over
the final configuration of the star weighted by rest-mass density at the 
end of the simulation. Thermal pressure generated by viscous heating 
causes $P/\rho^{\Gamma} > 1$ (recall $\kappa = 1$). We find that the 
viscous heating is much more significant in the low-density region than 
in the core.} \\
${}^{\rm c}$ {These values are obtained by solving 
Eqs.~(\ref{eq:jisco})--(\ref{eq:q}).} \\
${}^{\rm d}$ {The quantity $T/|W|$ is undefined when the star undergoes a
dynamical collapse. The number given here is an approximate value before 
the star becomes dynamically unstable.}
\end{minipage}
\label{tab:results}
\end{table*}

Four of the five stars we consider are hypermassive, and we expect
viscosity to change their structure significantly.
Star~I is the most hypermassive star ($M_0=1.38M_{\rm 0,sup}$, where
$M_{\rm 0,sup} = 0.221$ is the mass limit for uniformly rotating $n = 1$
polytropes, i.e.\ for stars at the mass-shedding 
limit~\cite{cst92}). We find that
this star eventually collapses to a black hole, but a substantial amount 
of rest mass is leftover to form a massive accretion disk. We consider 
three other hypermassive models (stars~II, III
and~IV) to study whether or not all hypermassive neutron stars will
collapse in the presence of viscosity.
Stars~II, III and~IV have the same rest mass ($M_0=1.13M_{\rm 0,sup}$) 
which is slightly smaller
than that of star~I but different
angular momenta $J$. We find that stars~III and~IV never
collapse, but evolve in a quasi-stationary manner to a uniformly 
rotating core surrounded by a low-density, disk-like envelope. Star~II eventually 
collapses to a black hole if we impose rapid cooling.  In the no-cooling
limit, however, this model forms a uniformly rotating core surrounded by a
substantial disk.
Star~V is the
only non-hypermassive model. As expected, this star does not collapse
under the action of viscosity. However, viscosity cannot drive the whole star 
to rigid rotation, because the angular momentum of the star exceeds 
the maximum angular momentum allowable for a rigidly-rotating star having 
the same rest mass. Instead, viscosity again leads to a uniformly 
rotating core and a differentially rotating disk-like envelope.
The final outcomes of the five models are summarized 
in Table~\ref{tab:results}.

We also performed 3D simulations on stars~I and~IV to search
for unstable, nonaxisymmetric secular modes. A nonaxisymmetric 
bar instability usually develops when a star is rotating rapidly, i.e.\ has a
sufficiently large $T/|W|$.  Of 
the five models we study, stars~I and~IV have the highest $T/|W|$. 
We do not find any nonaxisymmetric instabilities in these two models, 
and the 3D results roughly agree with the axisymmetric results. 

We discuss the results
of our simulations in detail in the following subsections.

\subsection{Star~I}
\subsubsection{2D with no cooling}
\begin{figure}
\vskip 1cm
\includegraphics[width=8cm]{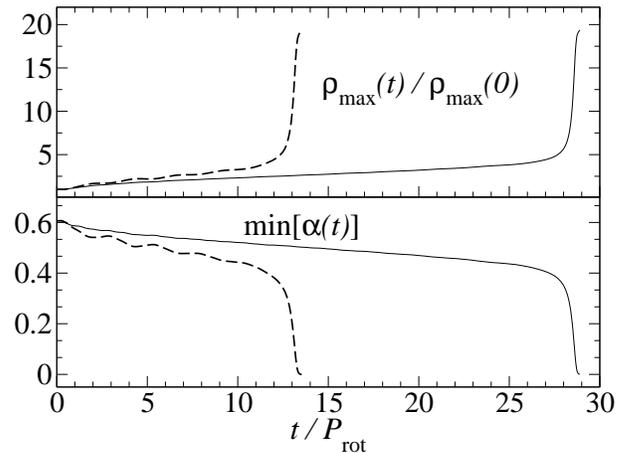}
\vskip 0.5cm
\caption{Maximum value of rest-mass density (upper panel) and 
minimum value of lapse (lower panel) as a function of time for 
star~I in the presence of viscosity. The solid (dashed) curves 
represent the case without (with) cooling. In both cases, the 
central core collapses to a black hole, and leaves behind a massive 
accretion disk.}
\label{fig:rho_alpha_I}
\end{figure}

\begin{figure*}
\includegraphics{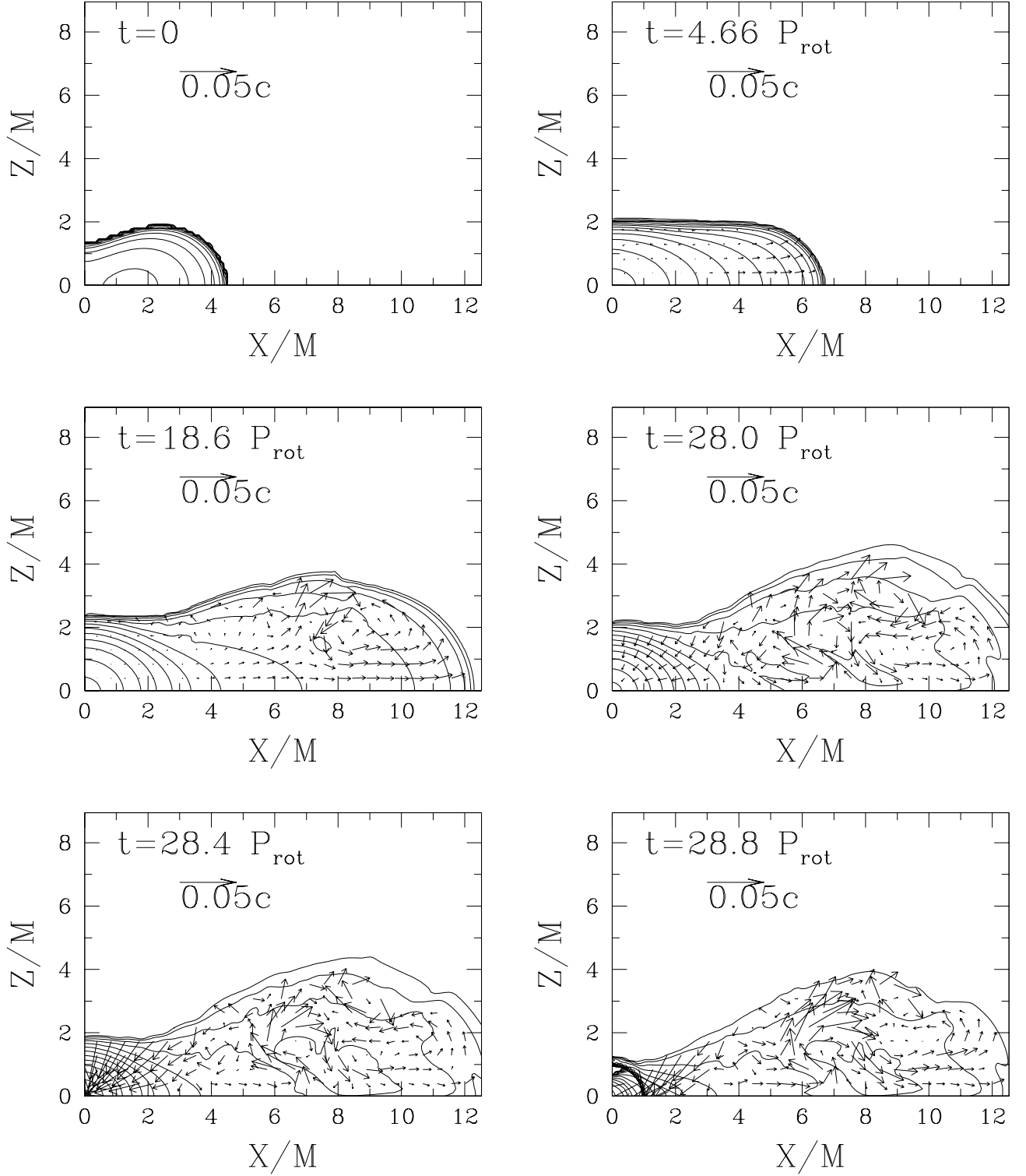}
\caption{Meridional rest-mass density contours and velocity field
at various times
for star~I. The simulation was performed by assuming that the system is
axisymmetric and experiences no cooling.
The levels of the contours (from inward to outward) are $\rho_0/\rho_{0,\rm max}
= 10^{-0.15 (2 j +0.6)}$, where $j=0,\ 1, \ ...\ 12$.
In the lower right panel ($t=28.8 P_{\rm rot}$), the thick curve
denotes the apparent horizon.}
\label{fig:contI}
\end{figure*}

Star~I is the most massive neutron star we study. 
We first perform an axisymmetric calculation with no cooling. At $t=0$, 
the star has a toroidal density profile, i.e., the maximum density occurs 
off center (see the upper left panel of Fig.~\ref{fig:contI}). As viscosity 
gradually brakes differential rotation, the star readjusts to a 
monotonic density profile. Figure~\ref{fig:rho_alpha_I} shows the maximum 
rest-mass density and the minimum value of the lapse as a function of time. 
Figure~\ref{fig:contI} 
shows the meridional rest-mass density contours at various times. We see that 
a meridional current is built up in the process. However, the 
magnitude of the meridional velocity ($\la 0.01c$) is much smaller than 
the typical rotational velocity ($\sim 0.3c$).

\begin{figure}
\includegraphics[width=8cm]{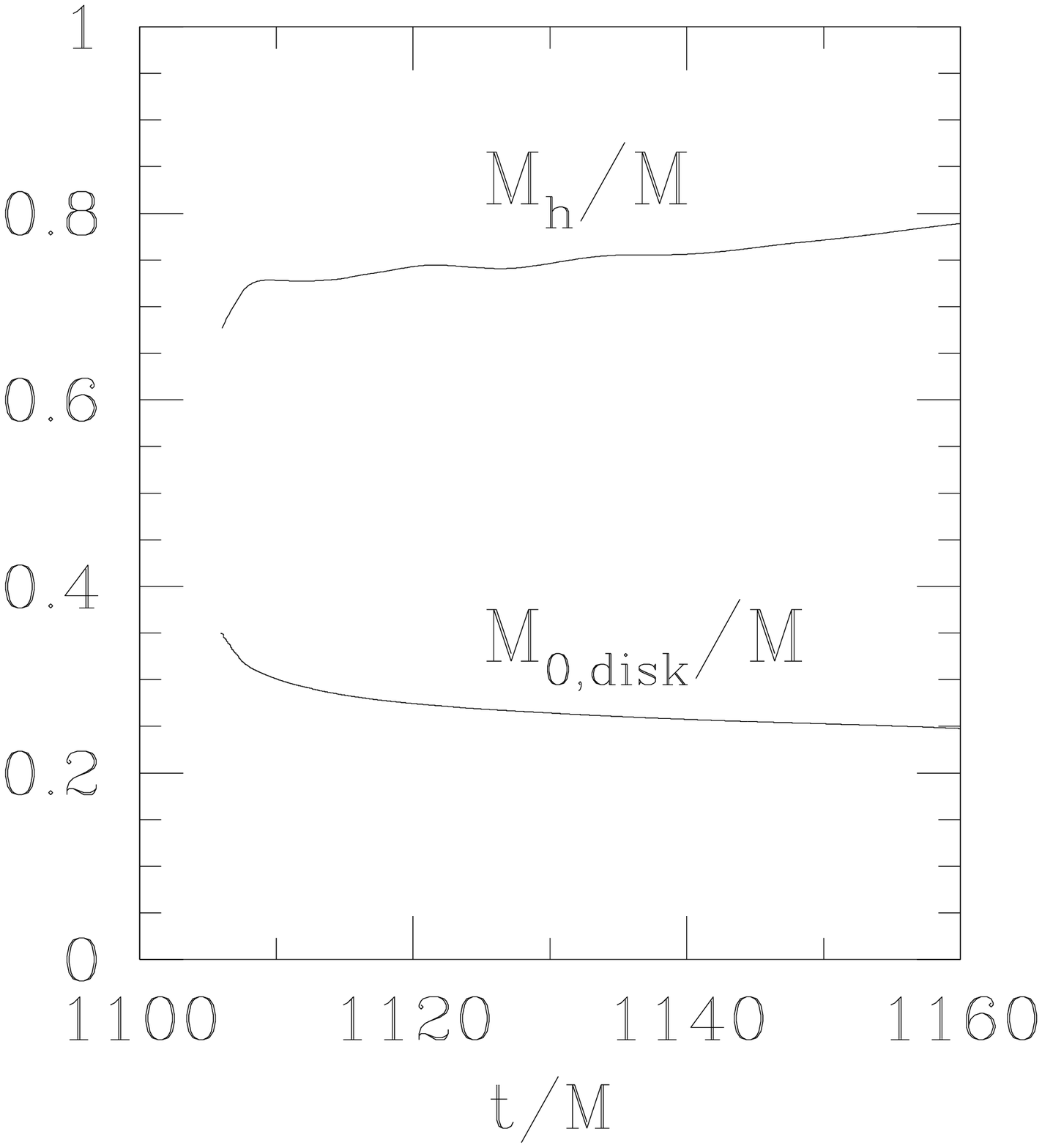}
\caption{Evolution of the black hole mass $M_h$ and the rest mass 
of the disk $M_{0,\rm disk}$ after the appearance of the apparent 
horizon at $t = 1106M=28.8P_{\rm rot}$. Note that time is plotted 
in units of $M$ ($1 P_{\rm rot}=38.4M$).  Black hole excision is
employed to track this late evolution.}
\label{fig:excision}
\end{figure}

Viscosity destroys differential rotation 
and transfers angular momentum to the outer layers. 
In the early phase of the evolution, the core contracts and 
the outer layers expand in a quasi-stationary manner. As the core 
becomes more and more rigidly-rotating, it approaches instability 
because the star is hypermassive and cannot 
support a massive rigidly-rotating core. At time 
$t \approx 27 P_{\rm rot} \approx 11 \tau_{\rm vis}$, 
the star becomes dynamically unstable and collapses. An apparent horizon 
appears at time $t \approx 28.8 P_{\rm rot}$. 
Without black hole excision, the code crashes about $10M$ after the
horizon appears because of grid stretching. 
About 30\% of rest mass remains outside the apparent horizon 
at this point.  We then continue the evolution 
using the excision technique described in~\cite{dsy03}. We are able 
to extend the evolution reliably for another $55M$. 
The system settles down to a 
black hole surrounded by a massive ambient disk. 
The rest mass $M_{0,\rm disk}$ and angular momentum $J_{\rm disk}$ 
of the disk can be calculated 
by integrating the rest-mass and angular momentum density over the
volume outside 
the apparent horizon [c.f.\ Eqs.~(\ref{def:M_0}) and~(\ref{Jaxi})]. 
The mass of the black hole can be estimated by the proper circumference 
of the horizon in the equatorial plane: 
$M_h = {\cal C}_h/4\pi$ (assuming that the spacetime 
can be described by a Kerr metric). Figure~\ref{fig:excision} shows 
the evolution of $M_h$ and the rest mass of the disk $M_{0,\rm disk}$. 
We estimate the final values of $M_h$, $M_{0,\rm disk}$, and
$J_{\rm disk}$ by fitting
these curves to analytic functions of the form $A + B \exp(-C t)$ and
extrapolating these fitting functions to $t \to \infty$~\cite{note:extrap}.
We estimate that the mass of the final black hole is $M_h \approx 0.82M$. 
The asymptotic rest mass and angular momentum of the ambient disk are
found to be
$M_{0,\rm disk} \approx 0.23 \ M_0$ and $J_{\rm disk} \approx 0.65 J$. 
We can infer from the conservation of angular momentum that the 
final angular momentum of the black hole is $J_h \approx 0.35J$. Hence we 
find $J_h/M_h^2 \approx 0.52 (J/M^2) \approx 0.52$.

The formation of a massive disk is mainly due to the fact that 
viscosity transports angular momentum from the inner core to 
the outer layers. The material in the outer region is unable to fall into 
the black hole because of the centrifugal barrier. The final mass 
of the black hole and disk can also be estimated independently
from the conservation of specific angular momentum using a method
developed by Shapiro and Shibata~\cite{ss02}, which we apply below.

During the dynamical collapse, the effect of viscosity is negligible. 
Since the spacetime is axisymmetric, the specific angular momentum 
$j=hu_{\varphi}$ of a fluid particle is conserved. For a Kerr black 
hole of mass $M_h$ and angular momentum $J_h$, the specific angular 
momentum of a particle at the innermost stable circular orbit (ISCO) 
$j_{\rm ISCO}$ is given by 
\begin{equation}
  j_{\rm ISCO} = \frac{ \sqrt{M_h r_{\rm ms}} (r_{\rm ms}^2 
-2a \sqrt{M_h r_{\rm ms}} + a^2)}{r_{\rm ms} (r_{\rm ms}^2
-3M_h r_{\rm ms} + 2a\sqrt{M_h r_{\rm ms}})^{1/2}} \ ,
\label{eq:jisco}
\end{equation}
where $a\equiv J_h/M_h$. The ISCO radius is 
\begin{equation}
  r_{\rm ms} =M_h [3+Z_2 - \sqrt{ (3-Z_1)(3+Z_1+2Z_2)} ] \ ,
\end{equation}
where 
\begin{equation}
  Z_1 = 1+\left(1-\frac{J_h^2}{M_h^4}\right)^{1/3} \left[ 
\left(1+\frac{J_h}{M_h^2}\right)^{1/3} + \left(1-\frac{J_h}
{M_h^2}\right)^{1/3} \right]  \ \ 
\end{equation}
and 
\begin{equation}
  Z_2 = \left( 3\frac{J_h^2}{M_h^4}+Z_1^2\right)^{1/2} \ .
\end{equation}
The rest mass and angular momentum of the escaping matter in 
the envelope with $j>j_{\rm ISCO}$ is given by 
\begin{eqnarray}
  M_{0,{\rm disk}} &=& \int_{j> j_{\rm ISCO}} \rho_* \, d^3 x \ , \label{eq:DeltaM0} \\ 
  J_{\rm disk} &=& \int_{j> j_{\rm ISCO}} \rho_* j \, d^3 x \ . \label{eq:DeltaJ}
\end{eqnarray}
We assume that the energy radiated by gravitational waves is negligible 
so that the total mass-energy of the system is approximately conserved. Hence we 
have 
\ba
  M &=& M_h +  M_{\rm disk} \ , \label{eq:Meq}\\
  J &=& J_h + J_{\rm disk} \ .  \label{eq:Jeq}
\ea
For a bound system, the contribution to the mass-energy of the matter 
in the disk, $M_{\rm disk}$, is smaller than its rest mass 
$M_{0,{\rm disk}}$. We write 
\begin{equation}
  M_{\rm disk} = q M_{0,{\rm disk}} \ .
\label{eq:q}
\end{equation}
We consider two opposite limits for $q$: $q=1$ and $q=M/M_0$. The 
value $q\approx 1$ is a good approximation in the weak gravity regime. 
In the limit where $M_h \ll M$, we have $M_{\rm disk} \approx M$ and 
$M_{0,{\rm disk}} \approx M_0$. Hence in this limit, $q \approx M/M_0$, 
which is 0.92 for star~I (see Table~\ref{tab:models}).  We expect that 
the correct $q$ lies somewhere between these two extremes, which are
not very different. 
The mass and angular momentum of the black hole can be estimated 
by solving the system of transcendental 
Eqs.~(\ref{eq:jisco})--(\ref{eq:q}) at a particular time slice during
the pre-excision dynamical collapse phase, including the time slice 
at the onset of dynamical collapse (where $q$ is close to unity). 
We find that the values of $M_h$ and $J_h$ are insensitive to
$q$. They are also insensitive to which time slice we choose to 
do the calculation. We find $M_h \approx 0.75 M$ and $J_h \approx 0.35 J$ 
($J_h/M_h^2 \approx 0.6$). The rest mass in the ambient disk is found to 
be $M_{0,{\rm disk}} \approx 0.23 M_0$. It should be noted 
that this calculation is based on the assumptions that the spacetime 
around the disk can be approximated by a Kerr metric, and that the disk 
is moderately thin and lies in the equatorial plane of the hole. 
This approximation is not reliable when the disk is massive
($M_{\rm disk}\approx M$). 
In our case, we find this calculation agrees rather well with the 
actual asymptotic values determined by the dynamical simulation with
excision.  Henceforth we will use 
Eqs.~(\ref{eq:jisco})--(\ref{eq:q}) to estimate $M_h$, $J_h$, 
$M_{0,\rm disk}$ and $J_{\rm disk}$ whenever an apparent horizon forms. 

\begin{figure}
\vskip 1cm
\includegraphics[width=8cm]{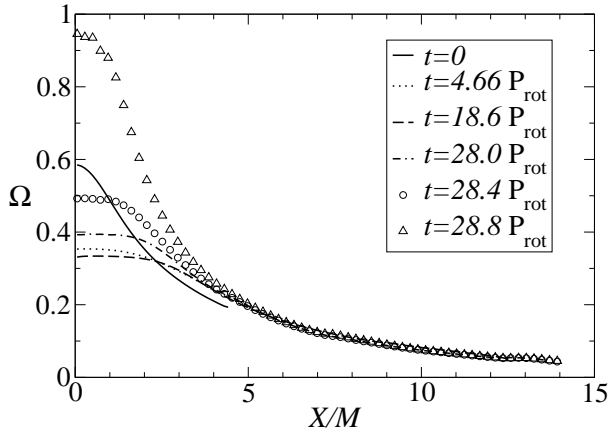}
\vskip 0.5cm
\caption{Angular velocity profiles in the equatorial plane at various
times during the evolution of star~I.}
\label{fig:OmegaI}
\end{figure}

Since viscosity is small in the low-density region, it takes  
longer to remove the differential rotation in the outer 
layers. Figure~\ref{fig:OmegaI} shows the angular velocity profiles 
at various times. We see that by the time the inner core collapses, 
the material in the outer layers is still differentially rotating. 
After the dynamical collapse, viscosity will cause 
some of the remaining material to slowly accrete onto the black hole.

We monitor the conserved quantities and the constraints during the 
entire evolution. Since our finite-difference scheme preserves
the rest mass, the variation of $M_0$ can only come from material 
flowing out of the grid. We find that $M_0$ is conserved 
to ~0.01\%, and angular momentum is conserved to ~0.1\%.
The Hamiltonian constraint is violated by less than 0.3\% 
before the dynamical collapse occurs. It increases to 3\% 
by the time an apparent horizon appears. The 
momentum constraints are violated by less than 1\% before the
dynamical collapse occurs, and increase to 6\% by the time 
an apparent horizon appears.

\subsubsection{2D with rapid cooling}
\label{2Dcooling}

\begin{figure*}
\includegraphics{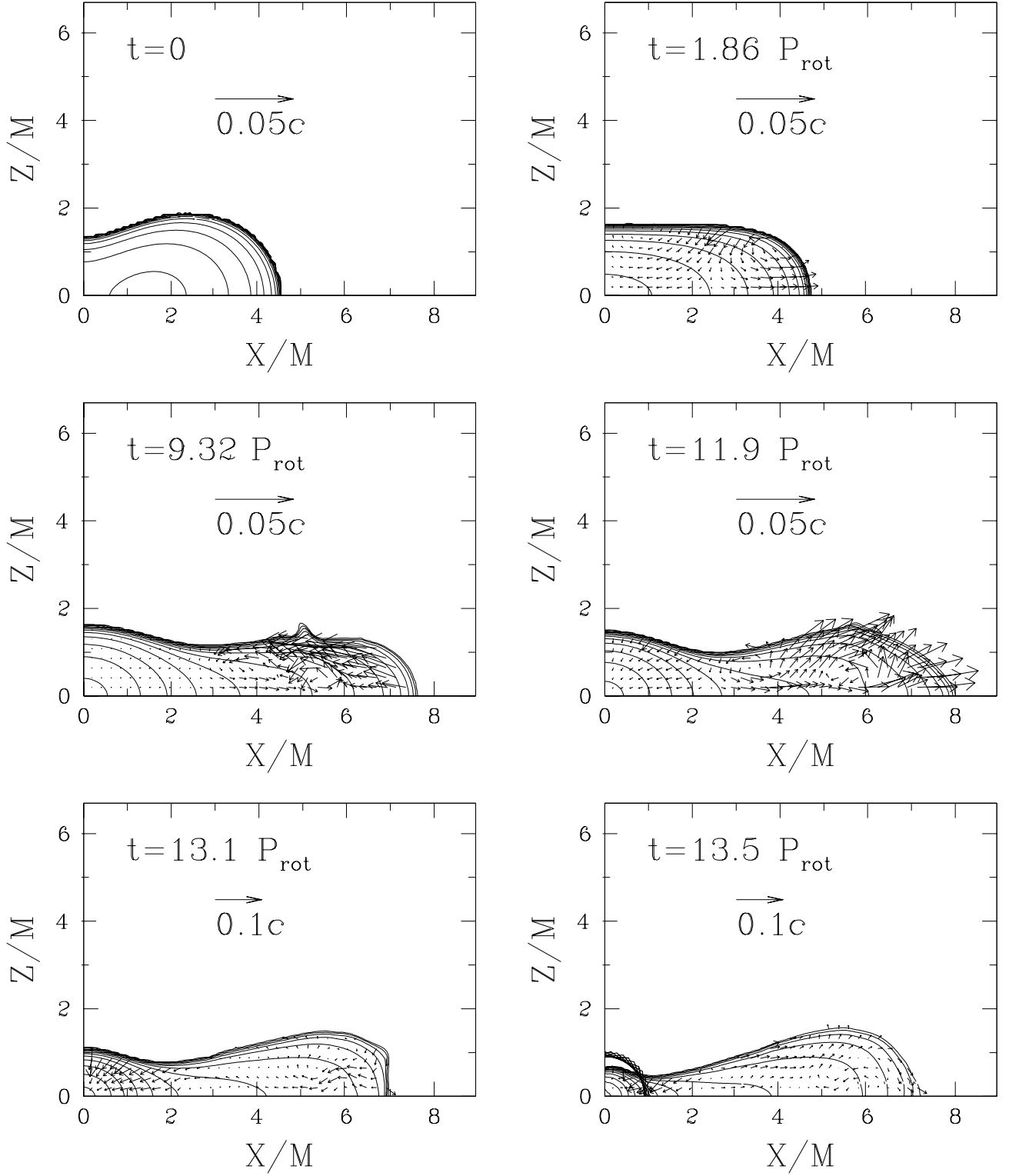}
\vskip 0.5cm
\caption{Same as Fig.~\ref{fig:contI} but for rapid cooling.}
\label{fig:contI_cool}
\end{figure*}

We next perform an axisymmetric simulation of star~I with rapid cooling. 
The dashed lines in Fig.~\ref{fig:rho_alpha_I} show the 
time evolution of the maximum 
rest-mass density and the minimum value of the lapse. As in 
the no-cooling case, the inner core contracts and the outer layers 
expand in a quasi-stationary manner. The core then collapses dynamically 
to a black hole and leaves behind a massive accretion disk. Since 
there is no viscous heating, the whole process occurs more quickly than
in the no-cooling case.  The dynamical collapse occurs at time 
$t\approx 12 P_{\rm rot} \approx 5 \tau_{\rm vis}$ 
and the apparent horizon appears at $t\approx 13.5 P_{\rm rot}$. 
Figure~\ref{fig:contI_cool} shows the meridional rest-mass density 
contours at various times. We estimate, by solving 
Eqs.~(\ref{eq:jisco})--(\ref{eq:q}) at $t \approx 13.5 P_{\rm rot}$ 
with $q=1$, that the mass and angular momentum of the final 
black hole are $M_h \approx 0.75 M$ and $J_h \approx 0.45 J$ 
($J_h/M_h^2 \approx 0.8$). About 20\% of rest mass escapes capture 
to form an accretion disk.

\begin{figure}
\vskip 1cm
\includegraphics[width=8cm]{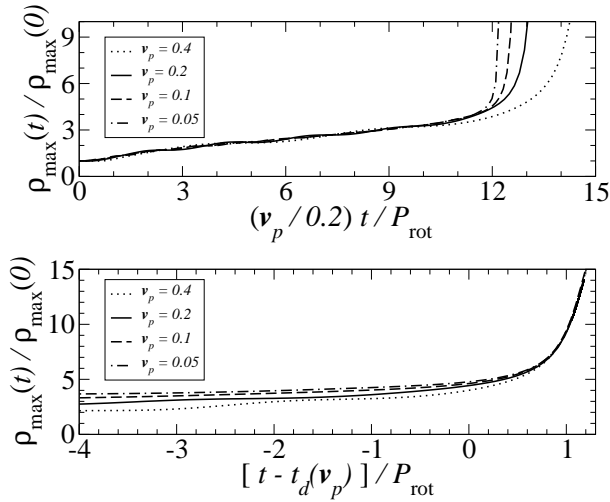}
\vskip 0.5cm
\caption{Evolution of the maximum rest-mass density of star~I for various 
strengths of viscosity, assuming rapid cooling.  Upper panel: the curves coincide when 
plotted against the scaled time prior to dynamical collapse. 
Lower panel: during dynamical collapse, it is possible to shift the time 
axes [$t \rightarrow t-t_d(\nu_P)$] so that the curves again coincide, which 
indicates that viscosity plays an insignificant role during the dynamical 
collapse phase.}
\label{fig:scaling}
\end{figure}

In Section~\ref{visctest}, we demonstrated that when the viscous timescale is 
significantly 
longer than the dynamical timescale, the secular rates of change of 
all physical 
quantities scale inversely with viscosity. The secular evolution of star~I with 
rapid cooling is short enough for us to perform another detailed scaling test. 
Figure~\ref{fig:scaling} demonstrates this scaling 
behavior by evolving star~I with four different 
strengths of viscosity $\nu_P=$ 0.4, 0.2, 0.1, and 0.05. 
(The curves in Fig.~\ref{fig:rho_alpha_I} correspond to $\nu_{\rm P} = 0.2$.)
We see that the scaling holds until dynamical collapse at time 
$t \approx t_d(\nu_P)$. When $t \gtrsim t_d$, the evolution 
of the system is no longer driven by viscosity. We therefore expect that 
the collapse is independent of the strength of viscosity as long as 
the viscous timescale is much longer than the dynamical timescale. 
In the lower panel of Fig.~\ref{fig:scaling}, we demonstrate that it 
is possible to 
shift the time axes ($t \rightarrow t-t_d$) so that the four 
viscosity runs yield the same result when $t - t_d \gtrsim 0$, which 
indicates that viscosity 
is insignificant during the dynamical collapse. The values 
of $t_d$ are determined by requiring that the scaling relation 
$t_d(\nu_2)/t_d(\nu_1) \approx \nu_1/\nu_2$ holds, and that the four curves
be aligned when plotted against the shifted time $t-t_d(\nu_P)$. 
We found that $t_d(\nu_P)/P_{\rm rot}\approx$ 6.1, 12.0, 24.08 and 47.75 
respectively for $\nu_P=$ 0.4, 0.2, 0.1, and 0.05. The fact that 
we are able to find $t_d(\nu_P)$ that satisfies these requirements 
validates our physical interpretation of the two phases of 
evolution.

\begin{figure}
\vskip 1cm
\includegraphics[width=8cm]{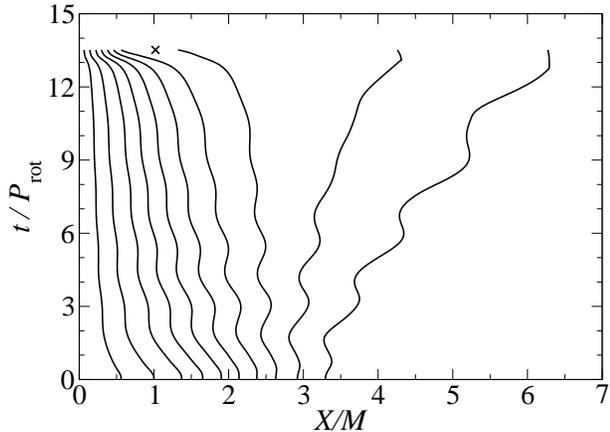}
\vskip 0.5cm
\caption{The worldlines of Lagrangian fluid elements at the equator
for star~I, assuming rapid cooling. 
The cylindrical coordinate $X$ of the particles at time 
$t=0$ is chosen so that the initial fraction of rest mass $m(X)$ 
interior to the cylinder of radius $X$ is,  
from left to right, $m$=0.03, 0.1, 0.2, 0.3, 0.4, 0.5, 0.6, 
0.7, 0.8 and 0.9. The cross in the diagram denotes the location 
of the apparent horizon at the end of the simulation.}
\label{fig:worldlines}
\end{figure}

To better visualize the effects of viscosity, we follow the motions 
of ten selected Lagrangian fluid elements. Figure~\ref{fig:worldlines}
shows the worldlines of these particles.
We choose the particles to be in the equatorial plane of the star. Equatorial 
symmetry implies that the particles will remain in the equatorial plane
at all times. The position of a fluid particle $X$ 
satisfies the equations
\begin{equation}
  \frac{d}{dt} X(t) = \frac{u^x(t;X(t))}{u^t(t;X(t))} \ \  . 
\end{equation}
We label the particles by the initial fraction 
of rest mass interior to the cylinder of radius $X$.
We see that the particles with the initial mass fraction $m \la m_* \simeq 0.8$ 
move toward the center and ultimately move inside the apparent horizon, 
while those with $m \ga m_*$ move away from the center and remain 
outside the apparent horizon. This agrees with our estimates 
of the rest mass of the ambient disk. 

\begin{figure}
\vskip 1cm
\includegraphics[width=8cm]{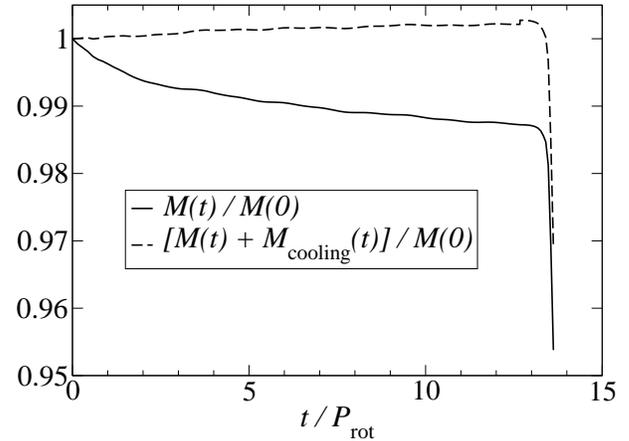}
\vskip 0.5cm
\caption{Evolution of the mass $M$ of star~I in the rapid-cooling limit. 
The mass is not conserved since the thermal energy generated by viscous 
heating is removed. However, the sum of the remaining mass and the mass 
carried away by ``radiation'', $M_{\rm cooling}$, is approximated conserved.} 
\label{fig:MI}
\end{figure}

Since there is rapid cooling, the mass is not 
conserved because the thermal energy generated by viscous heating is removed, 
as discussed in Section~\ref{diagnostics}. However, when we account 
for the mass-energy carried away by thermal radiation, $M_{\rm cooling}$, the 
total mass $M_{\rm tot}=M+M_{\rm cooling}$ should be conserved approximately
[see Eq.~(\ref{mtot})]. Figure~\ref{fig:MI} shows $M$ and 
$M_{\rm tot}$ as a function of time before an apparent horizon 
appears. The total mass is well-conserved except near the end 
of the simulation, where the numerical error arising from the grid 
stretching causes a few percent drop in the mass.

We monitored the violations of the constraints during the evolution. The 
violation of the Hamiltonian constraint is $\sim\!0.1\%$ before the 
dynamical collapse occurs, and goes up to 7\% at the time when the apparent 
horizon appears. The violations of the momentum constraints are 
also \mbox{$\sim\!0.1\%$} before the dynamical collapse occurs, and increase
to 8\% at the time when the apparent horizon appears.

\subsection{Other models}

\begin{figure}
\vskip 1.0cm
\includegraphics[width=8cm]{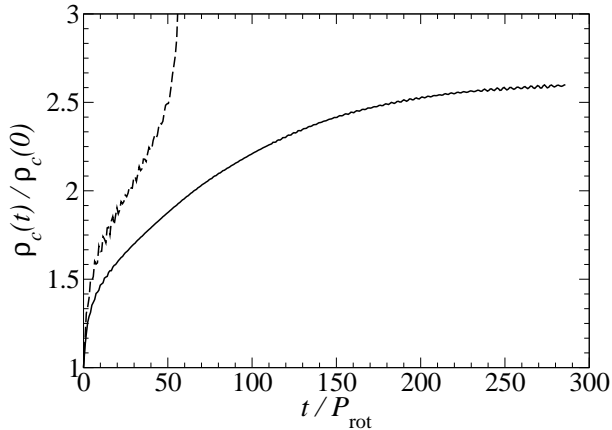}
\vskip 0.5cm
\caption{Evolution of the central rest-mass density for star~II 
with no cooling (solid line) and with rapid cooling (dashed line).}
\label{fig:cdenII}
\end{figure} 

The evolution of star~II due to viscosity is different 
from that of star~I. Although it is still hypermassive, the mass 
of star~II is smaller than that of star~I. When evolved in the absence of 
cooling, the star does not collapse to a black hole, but forms 
a rigidly-rotating core with a low-density disk-like envelope. 
When evolved in the rapid-cooling limit, however, the star collapses to a 
black hole. Figure~\ref{fig:cdenII} shows the evolution of the central 
density.

\vskip 0.5cm
\begin{figure}
\includegraphics{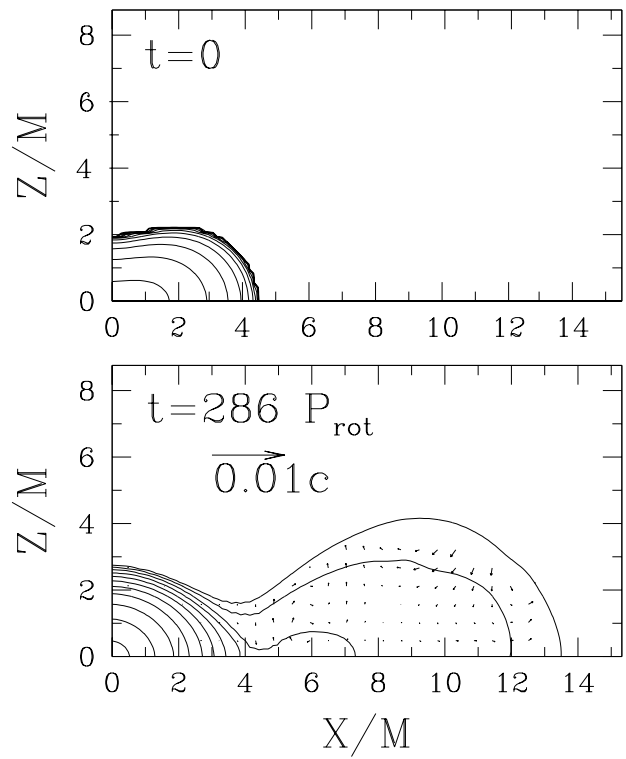}
\vskip 0.5cm
\caption{Meridional rest-mass density contours for star~II with no
cooling. The upper graph shows the contours at $t=0$ and the lower 
graph shows the contours at the end of the simulation 
($t=286 P_{\rm rot} = 87 \tau_{\rm vis}$). 
The contours are labeled as in Fig.~\ref{fig:contI}.}
\label{fig:contII}
\end{figure}

\begin{figure}
\vskip 1cm
\includegraphics[width=8cm]{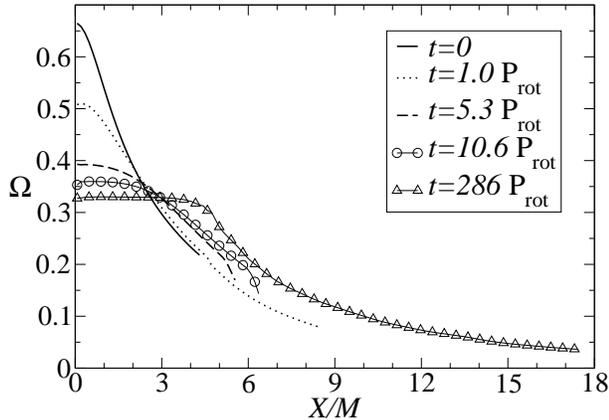}
\vskip 0.5cm
\caption{Angular velocity profiles at various times in 
the equatorial plane for star~II evolved with no cooling.}
\label{fig:Omega_II}
\end{figure}

\begin{figure}
\vskip 1cm
\includegraphics[width=8cm]{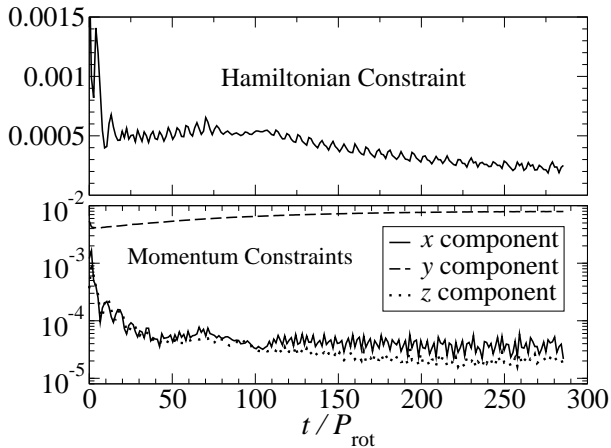}
\vskip 0.5cm
\caption{L2 norms of the Hamiltonian constraint and momentum 
constraints for star~II evolved without cooling.}
\label{fig:constraints}
\end{figure}

\begin{figure}
\vskip 1cm
\includegraphics[width=8cm]{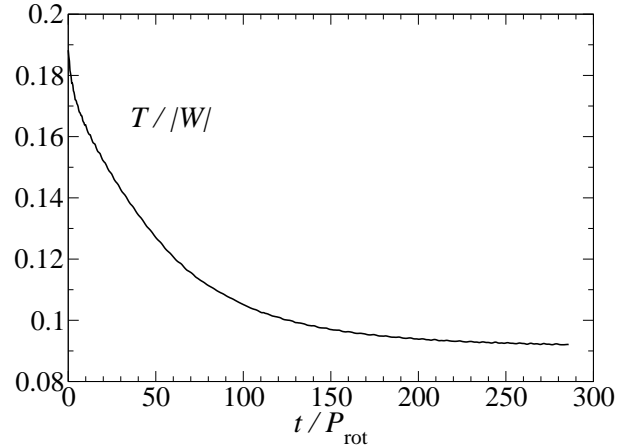}
\vskip 0.5cm
\caption{Evolution of $T/|W|$ for star~II evolved without cooling.}
\label{fig:ToWII}
\end{figure}

\begin{figure}
\vskip 1cm
\includegraphics[width=8cm]{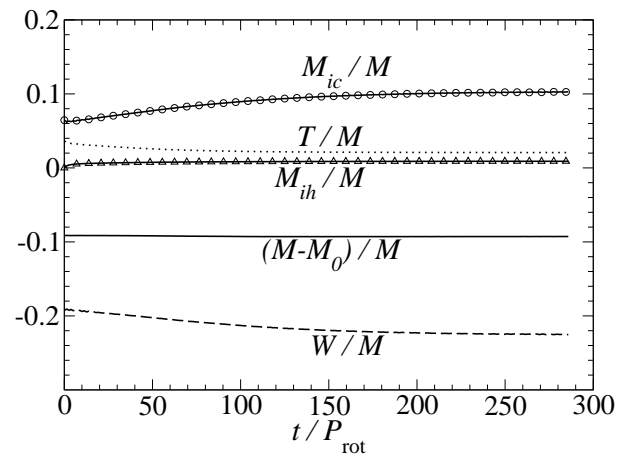}
\vskip 0.5cm
\caption{Evolution of various energies for star~II evolved with no cooling.}
\label{fig:energies}
\end{figure}

In the no-cooling case, Star~II has not collapsed to a black hole 
by the end of simulation ($t=286P_{\rm rot} = 87 \tau_{\rm vis} = 11,800 M$), 
but is settling to a uniformly rotating core surrounded by a massive torus.
Figure~\ref{fig:contII} 
shows the meridional density contours at the beginning and at the end 
of the simulation. Figure~\ref{fig:Omega_II} shows the angular velocity 
profiles at various times. Viscosity drives the star to a quasi-equilibrium,
rigidly-rotating core surrounded by a low-density 
disk. We cannot exclude the possibility that some of the 
outer material will slowly accrete onto the uniformly 
rotating inner core,  
eventually triggering collapse to a black hole. However the star 
acquires enhanced pressure support against 
collapse from viscous heating (i.e.\ $P/\rho_0^{\Gamma} > \kappa(0)$
where $\kappa(0) = 1$). Hence it may no
longer be hypermassive with respect to this new ``hot'' equation of 
state, as the simulation suggests. During the 
entire simulation, the star loses 1.2\% of its rest mass and 4.5\% of 
its angular momentum due to material flowing out of the grid. 
Figure~\ref{fig:constraints}
shows the L2 norms of the Hamiltonian and momentum constraints 
defined by Eqs.~(59) and~(60) of Paper I. We see that the 
violation of all the constraints are smaller than 1\% during the 
entire evolution of $286 P_{\rm rot}=11800~M$.

The values of  the ratio of kinetic to gravitational potential
energy, $T/|W|$, for all of the stars we studied decrease with time. 
Figure~\ref{fig:ToWII} shows the evolution of $T/|W|$ for star~II 
evolved without cooling. Viscosity transforms part of the rotational
kinetic energy into heat. 
It also changes the equilibrium configuration of the star
significantly, causing a redistribution of various energies.
Figure~\ref{fig:energies} 
shows the time evolution of various energies defined in 
Eqs.~(\ref{def:m0})-(\ref{mih}). 
The mass of the system 
decreases by 1.4\% due to a small amount of mass flowing out of the 
grid (not visible in the graph). The rotational 
kinetic energy $T$ decreases slightly. The contraction of the core 
raises the gravitational binding energy $|W|$, as well as 
the adiabatic part of the internal energy $M_{ic}$. Viscous heating 
generates the thermal energy $M_{ih}$, which prevents the star 
from undergoing catastrophic collapse.

In the rapid-cooling case, star~II collapses dynamically 
at time $t\approx 57 P_{\rm rot} \approx 17.4 \tau_{\rm vis}$. 
An apparent horizon appears at 
$t=57.7 P_{\rm rot}$. The mass and angular momentum of the 
final black hole are estimated by solving Eqs.~(\ref{eq:jisco})--(\ref{eq:q}):
$M_h \approx 0.88 M$ and $J_h \approx 0.63 J$ ($J_h/M_h^2 \approx 0.7$). 
About 10\% of rest mass is left as an accretion disk.

The situations for stars~III and~IV are similar. The inner core 
contracts in a quasi-stationary manner while the outer layers 
expand. Each system evolves into a rigidly-rotating core 
surrounded by a disk-like envelope. The stars do not collapse 
to black holes at the end of the simulations whether or not 
rapid cooling is imposed. Again, we do not rule out the 
possibility that they might collapse to black holes 
when enough material accretes onto the inner core.

From Table~\ref{tab:results}, we see that at the end of the simulation,
a large amount of angular momentum is transported to a massive disk.
For stars~III and~IV, the rest mass of the core $M_{0,\rm core}$ is
smaller than the rest-mass limit of a supramassive star $M_{0,\rm sup}$.
For star~II, $M_{0,\rm core}$ is slightly smaller than $M_{0,\rm sup}$
in the absence of cooling, but is slightly greater than $M_{0,\rm sup}$
in the rapid-cooling limit. For star~I, $M_{0,\rm core} > M_{0,\rm sup}$
in both the rapid-cooling and no-cooling cases. This suggests that
the fate of a hypermassive neutron star depends on whether
viscosity can create a rigidly-rotating core with
$M_{0,\rm core} > M_{0,\rm sup}$, in which case it will collapse. 
Both viscous heating and the star's initial angular momentum play
an important role in the final outcome. A hypermassive neutron
star with higher mass and lower angular momentum is prone to collapse,
whereas viscous heating tends to suppress the collapse.

Finally, we study the effect of viscosity on star~V, which is 
non-hypermassive. As expected, the star does not collapse to a 
black hole, irrespective of cooling. The star eventually 
evolves into a rigidly-rotating core surrounded by a disk. 
The fact that the star does not simply become rigidly 
rotating without shedding mass is due to the fact that viscosity 
conserves $M_0$ and $J$. For a given $M_0$ and equation of state, 
there is a maximum value of angular momentum 
$J_{\rm max}(M_0)$ above which a star can no longer maintain rigid 
rotation without shedding mass at the equator. In the case of
star~V, it is apparent that $J>J_{\rm max}$. Hence viscosity cannot 
force the whole star 
into rigid rotation. Similar results were found in studies 
of viscous evolution of differentially rotating white 
dwarfs assuming Newtonian gravitation~\cite{durisen}.

\subsection{3D tests of bar formation}
The results of the axisymmetric runs described above will not be 
physically relevant if the models are secularly unstable to bar formation. We 
evolved stars I and IV in 3D to check for the formation of bars.  
These models were chosen because, of our five models, they have the 
highest 
values of $T/|W|$.  Viscous heating can lead to expansion and, hence, 
a decrease in $T/|W|$. Thus, to further increase the chance of bar 
formation, we performed these runs in the rapid-cooling limit.  
We superimposed $m=2$ perturbations on the initial data according to
Eq.~(\ref{seedbar}) for both stars with magnitude $\delta_b~=~0.1$, 
so that $|Q| = 0.014$ initially.  Both runs were performed in 
$\pi$-symmetry with a uniform grid of size $128\times64\times32$ and outer 
boundaries in the $x$-$y$ equatorial plane at $14.3 M$ for star I and
$17.1 M$ for star IV. 
To reduce computational costs, the extent of the grid in the 
$z$-direction is only half as large as in the $x$- and $y$-directions.
This setup is feasible because these 
models are initially highly flattened due to rapid rotation and because 
their evolution results in an expansion which is largely horizontal.  
For star I, we find that $|Q|$ decreases 
in magnitude until the code terminates due to the collapse, when 
$|Q| = 0.0015$.  Before the collapse, all constraints are 
satisfied to within $3.5\%$ while $M$ and $J$ are conserved to within 
$3\%$.  For star IV, $|Q|$ also decreases, reaching 0.0023 after 
$ 11.5 P_{\rm rot} = 3.3 \tau_{\rm vis}$, when the simulation is
terminated.  In this case, the constraints 
are satisfied to within $6.0\%$ while $M$ and $J$ are conserved to 
within $1.4\%$. We also find that the rest density contours 
remain nearly axisymmetric throughout the evolutions of both stars. 
These results indicate that both stars are stable 
against secular bar formation on the viscous timescale. 

\section{Discussion and Conclusions}

We have simulated the evolution of rapidly rotating stars in full
general relativity including, for the first time, shear viscosity.
Our findings indicate that the braking of differential rotation in 
hypermassive stars always leads to significant structural changes,
and often to delayed gravitational collapse.  The rest 
mass, angular momentum, and thermal energy all play a role
in determining the final state.  
We performed axisymmetric numerical simulations of five models to 
study the influence of these parameters. 
In the presence of shear viscosity,
the most hypermassive model which we studied (star~I), collapses to 
a black hole whether we evolve by ignoring cooling, or by assuming
rapid cooling of the thermal energy generated by viscosity.  However, the 
viscous transport of
angular momentum to the outer layers of the star results in mass 
outflow and the formation of an appreciable disk.  Next, we considered 
three hypermassive models (stars~II, III, and~IV) with the same 
rest mass $M_0$, but different values of the spin parameter $J/M^2$.  
These models have smaller $M_0$ than star~I, 
and are therefore less prone to collapse.  Star~II, which has
$J/M^2 = 0.85$, collapses when evolved in 
the rapid-cooling limit, leaving behind a disk.  But without 
cooling, this model evolves to a stable, uniformly rotating core with
a differentially rotating massive disk.  The additional thermal 
pressure support provided by viscous heating prevents collapse 
in this case.     

In contrast, stars~III and~IV, which have $J/M^2 = 1.0$ and 1.1, 
respectively, do not collapse even in the rapid-cooling limit.
This is sensible because these models have a smaller rest mass than
star~I, but larger angular momenta than star~II.  Though the cores of
stars~III and~IV contract, they 
are prevented by centrifugal support from reaching the necessary 
compaction to become dynamically unstable.  
In both cases, we find low-density disks surrounding uniformly rotating 
cores.  However, our simulations do not rule out the possibility that 
slow accretion of the disk material could eventually drive the uniformly
rotating cores to collapse. 
Disk formation also occurs for star~V, which is differentially rotating
but non-hypermassive. Since there exist
stable, uniformly rotating models with the 
same rest mass, the braking of differential rotation in this case
does not result in collapse.  However, differentially rotating stars
can support larger $T/|W|$ than uniformly rotating stars.  In the
case of star~V, there does not exist a uniformly rotating star with
the same (high) angular momentum and rest mass, so that mass shedding must 
take place as
viscosity drives the star to uniform rotation.  In the final state,
we find a rigidly-rotating core surrounded by a low-density, disk.

Since results obtained from our axisymmetric code are physically reliable
only for models which are not subject to nonaxisymmetric instabilities,
we evolved stars~I and~IV in 3D to check for such instabilities.  Previous 
studies in Newtonian gravity 
have found that the secular, viscosity-driven bar instability
in uniformly rotating stars
should set in when $T/|W| \gtrsim 0.14$ 
\cite{c69,sl96}.  When general relativity is taken into account, 
the threshold value can be 
somewhat higher \cite{sz98}. Thus, of all of our models,
stars~I and~IV have the best chances of developing bars since they have
the highest $T/|W|$.  We introduced 
an initial bar-shaped perturbation and ran these cases in the 
rapid-cooling limit.  We found that, in both cases, the small initial 
perturbation
decays and no bar is formed.  This is somewhat surprising
since $T/|W|$ is well above 0.14 in both of these cases.
We plan to address this issue in a future report.  
%We note, however, that the viscosity-driven bar instability has
%only been studied in uniformly rotating stars 
%\cite{c69,j64,sl96}.  These results may not
%apply to strongly differentially rotating stars.  

For the evolution of each of our  five models, we find that a massive 
disk or torus forms in the final state. The disk typically carries 
$\sim\!20\%$ of the rest mass of the initial configuration.   
Viscosity transports angular momentum from the interior
of the star to the more slowly rotating exterior.  The exterior 
regions then expand to accommodate the additional centrifugal force, 
forming a low-density disk.  The inner core becomes rigidly 
rotating and,
in some cases, undergoes gravitational collapse. The disk, however,
remains differentially rotating since viscosity acts much more 
slowly in low-density regions.  
For cases in which black holes are formed,
the mass of the disk may be estimated by 
integrating the rest-mass density for those fluid elements
which have specific angular momentum $j$
greater than the value at the ISCO, $j_{\rm ISCO}$ [see 
Eq.~(\ref{eq:DeltaM0})].  The 
estimates obtained in this way agree reasonably well with the
results of our numerical simulations.  
Particularly good agreement was found for the case of star~I 
with no cooling, for which we were able to extend the evolution
some $55 M$ beyond the first appearance of an apparent horizon. 
The rest mass and angular momentum of the disk surrounding the 
rotating black hole could then be calculated directly and agreed
well with the estimates.  We expect that excision techniques will
continue to be crucial in establishing the final fate of 
systems involving matter surrounding black holes.  

%To firmly establish the 
%final fate of such a system with a disk surrounding a central black 
%hole, black hole excision is necessary. In a forthcoming paper 
%\cite{dsy03} a method for excising the singularity in a hydrodynamic 
%matter field is described.  
%Using our current implementation
%at the resolutions of this paper, 
%however, provides accurate evolutions of rotating systems only for times 
%$\sim 50 M$ following the excision of the black hole.  Since 
%longer evolutions are required to probe the final state, applying
%our excision code in the current analysis is not useful.

In a recent paper, Shibata \cite{s03} numerically simulated 
collapses of marginally stable, supramassive stars.  
These supramassive models were constructed using polytropic 
equations of state  with
$2/3 < n < 2$ and rotate at the mass-shedding limit with 
$0.388 \leq J/M^2 \leq 0.670$.  Shibata found that the collapse 
of these stars results in Kerr black holes and that no more than $0.1\%$ of the 
initial rest mass remains outside of the hole.  This result
is quite different from our finding that disks are usually present
following collapse.
However, the initial data for the two calculations are quite different,
as well as our inclusion of viscosity.
The analysis of \cite{s03} takes uniformly rotating, unstable 
configurations as initial data and follows their dynamical 
evolution. Our calculations begin with 
differentially rotating, stable configurations and follow 
both their secular (viscous) and dynamical evolution. Viscosity
drives our configurations to uniform rotation.  
We find that massive disks usually form as by-products of the 
formation of uniformly rotating cores.  This is due primarily 
to the transport of angular 
momentum from the inner to the outer layers.  In addition, all of our 
models have $0.85 \leq J/M^2 \leq 1.1$.  (Large angular momentum 
is required to generate a hypermassive neutron star in equilibrium.) 
Since this range is higher than that considered in \cite{s03},
our models more naturally produce disks \cite{note:disks}.

All of the phenomena observed in our simulations follow from
the braking of differential rotation in strongly relativistic
stars.  This may be accomplished by viscosity as shown here, but 
magnetic fields are likely to be more important.  The fate of 
the hypermassive remnants of binary neutron star mergers
may crucially depend on these effects.  The loss of
differential rotation support in such a remnant may lead to 
delayed gravitational collapse.  This collapse could in turn 
lead
to a delayed gravitational wave burst following the quasi-periodic
inspiral and merger signal~\cite{bss00}.  Our results indicate that if the 
remnant is not sufficiently hypermassive, collapse 
may not occur, at least not until the star cools by radiating
away its thermal energy.  Understanding the evolution of such merger 
remnants could aid the interpretation of signals observed by 
ground based gravitational wave detectors, such as LIGO, VIRGO,
GEO, and TAMA.  In addition, short-duration GRBs are thought
to result from mergers of binary neutron stars or neutron 
star-black hole systems \cite{rosswog,npk01}.  In this scenario, 
the GRB may be powered by accretion from a massive torus or disk 
surrounding a rotating black hole.  We have demonstrated that 
such disks are easily produced during the evolution of 
hypermassive neutron stars. 

The braking of differential rotation may also be important in
neutron stars formed in core collapse supernovae.  
Nascent neutron stars are probably characterized by significant 
differential rotation (see, e.g., \cite{zm97,rmr98,ll01,l02}
and references therein).  Conservation of angular momentum during
the collapse is expected to result in a large value of $T/|W|$.
However, uniform rotation can only support small values of $T/|W|$ 
without shedding mass (\cite{st83}, Chap. 7).  Thus, nascent
neutron stars from supernovae probably rotate differentially.
If the induced differential rotation
is strong enough, hypermassive neutron stars can form.
Their subsequent evolution and final fate then depends on 
the presence of viscosity or magnetic fields.  
Such considerations may be important for long-duration GRBs
in the collapsar model \cite{mw99}.  In this
model, the GRB is powered by accretion onto the central black
hole formed through core collapse in a massive star.

Several interesting astrophysical systems undergo secular evolution
in strongly gravitating environments.  In this paper,
we have shown that it is possible to study secular effects that occur 
over many dynamical timescales using hydrodynamic computations in full
general relativity.  We consider this an important step toward 
future numerical explorations of secular effects in other contexts.  
In particular,
we plan to incorporate MHD into our evolution code, as
magnetic braking probably acts more quickly than viscosity to 
destroy differential rotation in many systems, like neutron stars or 
supermassive stars \cite{smspapers}.  Our results have
also raised the following interesting question:  Under what 
circumstances are differentially rotating, compressible neutron 
stars with high $T/|W|$ unstable to nonaxisymmetric modes?  We 
plan to address this issue in a future report.

\acknowledgments

The calculations for this paper were performed
at the National Center for Supercomputing Applications at the
University of Illinois at Urbana-Champaign (UIUC).  This paper was
supported in part by NSF Grants PHY-0090310 and PHY-0205155 and
NASA Grant NAG 5-10781 at UIUC.

\begin{appendix}

\section{Viscous heating and radiative cooling}
\label{cooling_app}

In this appendix, we describe the thermal properties of our configurations.
The dissipation of rotational energy by viscosity heats the stars, 
but they may be cooled by radiation (e.g. neutrino radiation).  The presence
of radiation contributes a term $R^{\alpha\beta}$ to the stress tensor: 
$T^{\mu\nu}_{\rm total} = T^{\mu\nu} + R^{\mu\nu}$, with $T^{\mu\nu}$
given by Eq.~(\ref{Tuv}).  This modifies the equations of motion 
(\ref{eom}) to
\begin{equation}
\label{T_cooling}
T^{\mu\nu}{}_{;\nu} = - R^{\mu\nu}{}_{;\nu} = G^{\mu} \ ,
\end{equation}
where $G^{\mu}$ is the 4-force density due to radiation (see \cite{mm99}).
The specific entropy $s$ of a fluid element with temperature
$T$ and number density $n = \rho_0/m$ changes when there is
heating and cooling according to
\begin{equation}
  \label{dsdtau}
  nT{ds\over d\tau} = \Gamma_{\rm heat} - \Lambda\ = \Gamma_{\rm vis} + 
\Gamma_{\rm rad} - \Lambda_{\rm rad} \ ,
\end{equation}
where $\tau$ is the proper time along the element's worldline. Here,
we have separated the contributions from viscosity and 
radiation to the heating rate.  The quantity $\Gamma_{\rm vis}$ 
is the viscous heating rate per unit volume, which comes from the
$\sigma_{\alpha\beta}\sigma^{\alpha\beta}$ term in Eq.~(\ref{evolve_estar}). 
In terms of the thermal energy density $U_{\rm therm}$ and viscous 
timescale $\tau_{\rm vis}$, $\Gamma_{\rm vis}$ is roughly
\begin{equation}
\Gamma_{\rm vis} \simeq U_{\rm therm}/\tau_{\rm vis} \ .
\end{equation}
In general, a fluid can be heated and cooled by the presence of a radiation 
field.  The energy equation becomes
\begin{equation}
u_{\mu}T^{\mu\nu}{}_{;\nu} =  u_{\nu}G^{\nu} = -G^{\hat{0}} \ ,
\end{equation}
where the last equality arises from evaluating $u_{\mu}G^{\mu}$ in 
the comoving orthonormal frame of the fluid, where $u^{\hat{\mu}}=(1,0,0,0)$.  Then
\begin{equation}
\label{G0eqn}
G^{\hat{0}} = \Gamma_{\rm rad} - \Lambda_{\rm rad} = 
\int\!\!\int d\nu\,d\Omega (\kappa_{\nu}I_{\nu} - \eta_{\nu}) \ ,
\end{equation}
where the integral is evaluated in the comoving frame and $\kappa_{\nu}$, 
$I_{\nu}$, and $\eta_{\nu}$ are the opacity, intensity, and 
emissivity at frequency $\nu$ \cite{mm99}.  For applications of interest
here, radiation mediates {\it net} cooling of the viscous-heated fluid. 
Hence, we can set $\Gamma_{\rm rad} = 0$ for simplicity.

The first law of thermodynamics
\begin{equation}
d(e/n) = -Pd(1/n) + Tds \ , \qquad e \equiv \rho_0(1 + \epsilon)
\end{equation}
and Eq.~(\ref{ideal_P}) give
\begin{eqnarray}
\label{ds}
nTds &=& n^{\Gamma}d\left({Pn^{-\Gamma}\over\Gamma-1}\right) \nonumber \\
     &=& {n^{\Gamma}\over\Gamma-1}d\kappa \ .
\end{eqnarray}
Here we define the entropy parameter $\kappa$ by
$P \equiv \kappa n^{\Gamma}$, where, in general, $\kappa = \kappa(s)$. 
The form of $\Lambda_{\rm rad}$ will depend
on the details of the neutron star's microphysics, but it must
have the property that $\Lambda_{\rm rad} = 0$ when 
$\kappa(s) = \kappa_0 = \kappa(s=0)$,
where $\kappa_0$ is the value of $\kappa$ for the unheated fluid
(i.e. no emission from a zero-entropy fluid). 
Accordingly, we replace Eq.~(\ref{G0eqn}) for $\Lambda_{\rm rad}$ 
by the following illustrative form
\begin{equation}
\label{cooling_rate}
\Lambda_{\rm rad} = \xi n [\kappa(s)-\kappa_0]/\tau_{\rm cool}\ ,
\end{equation}
where $\xi$ is a constant and $\tau_{\rm cool}$ is the radiation
timescale.  Combining equations (\ref{dsdtau})-(\ref{cooling_rate}),
we find
\begin{equation}
  {d\kappa\over d\tau} = {\Gamma - 1\over n^{\Gamma}}
  \left\{\frac{U_{\rm therm}}{\tau_{\rm vis}} -
    \frac{\xi n[\kappa(s)-\kappa_0]}{\tau_{\rm cool}}\right\} \ .
\label{evolve_kappa}
\end{equation}
In the limit $\tau_{\rm cool} \gg \tau_{\rm vis}$, radiative cooling is 
unimportant and $\kappa$ increases due to viscous heating.  We refer to 
this regime as the {\em no-cooling} limit.
If $\tau_{\rm cool} \ll \tau_{\rm vis}$, then the first term on 
the right hand side of Eq.~(\ref{evolve_kappa}) may be dropped in 
relation to the second, giving
\begin{equation}
\frac{d}{d\tau}(\kappa - \kappa_0) = -\frac{\xi (\Gamma-1)}{n^{\Gamma-1}} 
\frac{(\kappa - \kappa_0)}{\tau_{\rm cool}} \ .
\end{equation}
Thus, $\kappa$ is exponentially driven to $\kappa_0$.  We refer to this 
regime as the {\em rapid-cooling} limit, whereby the thermal energy
generated by viscosity is radiated immediately and does not heat the gas. 
In effect, $\Lambda_{\rm rad} = \Gamma_{\rm vis}$ in this limit. 
In practice we implement this limit by omitting the 
$\sigma_{\mu\nu}\sigma^{\mu\nu}$ in Eq.~(\ref{evolve_estar}). 
Though we consider only these two limits in our analysis, we expect 
that, in reality, heating will dominate in some regimes and cooling in
others.  We treat both limiting cases in our simulations.

\end{appendix}


\begin{thebibliography}{99}
\bibitem{footnote0} A similar approach has been suggested
in~\cite{rsabs99} to simulate $r$-mode evolution and has been successfully
applied in~\cite{ltv01}.
\bibitem{note:modelsAB} In Paper~I, star~A was labeled ``C'' and star~B
         was labeled ``D''.
\bibitem{note:extrap}
  What appears to be a drift in $M_h$ due to numerical error becomes
  visible towards the end of the simulation.  This makes the
  extrapolated value of $M_h$ less reliable than the values of
  $M_{0,\rm disk}$ and $J_{\rm disk}$.
\end{thebibliography}
\end{document}